\documentclass[aps, prd,twocolumn, superscriptaddress, showpacs, nofootinbib]{revtex4-2}
\usepackage[utf8]{inputenc}
\usepackage[T1]{fontenc}
\usepackage{ae,aecompl} 
\usepackage{graphicx}
\usepackage{amsmath}
\usepackage{color}
\usepackage{amssymb}
\usepackage{latexsym}
\usepackage{wasysym}
\usepackage{psfrag}
\usepackage{comment}
\usepackage{ifthen}
\usepackage{hyperref}
\usepackage{longtable}
\usepackage{float}
\usepackage[utf8]{inputenc}
\usepackage{lineno}
\usepackage{units}
\usepackage{multirow}
\usepackage{orcidlink}
\usepackage{listings}
\usepackage{xcolor}
\usepackage{enumitem}
\usepackage{tikz}
\begin{abstract}
The Einstein Telescope (ET), a planned third-generation gravitational-wave (GW) observatory, will offer significantly improved sensitivity, introducing new challenges for data analysis and computing. To prepare for these demands, the ET community has initiated a series of Mock Data Challenges (MDCs) aimed at developing and testing analysis pipelines under realistic conditions.
This paper presents the first ET MDC, providing an overview of the simulated dataset and the properties of the injected GW signals, with a focus on populations of compact binary coalescences and Gaussian noise. A tutorial is also included to guide users in accessing the data and performing basic analyses.
This initial challenge establishes a baseline for future MDCs and supports collaborative efforts toward the successful scientific operation of the ET.
\end{abstract}

\begin{document}
\title{A mock data challenge for next-generation detectors}

\author{Tania Regimbau\, \orcidlink{0000-0002-0631-1198}}
\email{regimbau@lapp.in2p3.fr}
\affiliation{Laboratoire d'Annecy de Physique des Particules, CNRS, 9 Chemin de Bellevue, 74941 Annecy, France}

\author{Jishnu Suresh\, \orcidlink{0000-0003-2389-6666}}
\affiliation{Universit\'e C$\hat{o}$te d’Azur, Observatoire de la C$\hat{o}$te d’Azur, CNRS, Artemis, 06304 Nice, France}
\maketitle
\date{\today}

\section{Introduction}

Gravitational wave (GW) astronomy began with the historic first direct detection of a GW signal on September 14, 2015, when the second-generation of the LIGO detectors commenced operations~\cite{LIGOScientific:2016aoc}. The signal—the merger of two stellar-mass black holes—was identified by multiple data analysis pipelines, including two pipelines based on matched filtering, and a pipeline designed for unmodeled sources.

The development of these pipelines started decades earlier, in the 1990s, and involved extensive testing and optimization on simulated data, also called mock data challenges (MDCs). Initially, these tests were conducted on simulated noise, even before the first generation of detectors was built. Later, as LIGO and Virgo came online, simulated signals were injected into real detector noise, providing crucial insights into the performance and reliability of the detection methods.

Today, the field is rapidly progressing toward a new era of GW observatories. Proposed third-generation detectors, including the Einstein Telescope (ET) and the Cosmic Explorer (CE), are expected to be ten times more sensitive than the current generation, with access to lower frequency bands down to 1–5 Hz. This increased sensitivity and expanded frequency range will enable the detection of a much larger and more diverse population of sources, but it will also introduce new challenges. In particular, these detectors are expected to operate in a signal-rich regime, similar to the upcoming space-based mission LISA, where numerous long-duration and overlapping signals will simultaneously be present in the data.

In anticipation of these challenges, the ET community has initiated a new series of mock data challenges to support the development and validation of data analysis methods suited for the ET era. This effort builds on a first series of ET MDCs conducted between 2012 and 2016 \cite{2012PhRvD..86l2001R,2014PhRvD..89h4046R,2016PhRvD..93b4018M,2015PhRvD..92f3002M}, which demonstrated the feasibility of extending the analysis down to 5 Hz and introduced innovative techniques such as the use of the null stream in the ET’s triangular configuration to reconstruct the noise power spectral density.

Mock data challenges are critical in GW astronomy. They are essential for validating detection techniques, and ensuring that the pipelines used to identify GWs perform accurately and reliably. MDCs are also crucial for developing and refining parameter estimation algorithms, which are key to extracting astrophysical information. For instance, in addition to ground-based detectors like LIGO, Virgo, and KAGRA, the LISA mission has extensively used MDCs to develop methods for analyzing its signal-rich environment \cite{LISAMDC,2007CQGra..24S.551A,2022arXiv220412142B}. This work led to the development of global fit techniques capable of modeling and extracting multiple overlapping sources simultaneously. MDCs are also important for testing the performance of different detector configurations and verifying that computing infrastructures can handle the large data volumes and processing demands expected in the coming decades.

This paper presents the first round of the new ET mock data challenges, named \textit{ET-MDC1}. As a first step, \textit{ET-MDC1} adopts simplified models for both detector noise and the source population, focusing exclusively on compact binary coalescence signals in stationary Gaussian noise. This controlled setup establishes a clear baseline for pipeline validation, while increasing realism will be introduced in future mock data challenges. Section 2 describes the methodology used to generate the dataset, including the Monte Carlo simulations employed. Section 3 explores the statistical properties of the GW signals, detailing the characteristics of the simulated source population. Section 4 outlines the challenges addressed in this round of MDCs: the beginner challenge, which focuses on detecting the six loudest events in the dataset, and the expert challenge, which requires analyzing the full dataset and accounting for long-duration and overlapping sources. Section 5 provides instructions on how to access the dataset. Section 6 provides a simple tutorial to help get started with the MDC dataset. Finally, the conclusion discusses plans for future mock data challenges and outlines directions for advancing data analysis techniques in preparation for the ET.

\section{Description of the Simulation}

ET-MDC1 consists of time series data containing both instrumental noise and gravitational wave signals from compact binary coalescences (CBCs). Throughout this paper, we denote by $h$ the gravitational-wave signal
and by $s$ the detector strain data, which include both the GW signal
and instrumental noise.
The dataset covers approximately one month and is split into 2048-second segments to manage the computational load. The data are sampled at 8192 Hz, which provides sufficient resolution to capture the full GW signal from CBCs. The low-frequency cutoff is set at 5 Hz, consistent with the ET's design sensitivity. The simulation code is available on the ET GitLab repository\footnote{\url{https://gitlab.et-gw.eu/osb/div10/mdc-generation}}.

\subsection{Detector network}

For this first round, we focus solely on the triangular configuration of the ET, consisting of three independent V-shaped Michelson interferometers, each with 60-degree opening angles and 10\,km-long arms (denoted as E1, E2, and E3). A subsequent design modification, known as the ``xylophone'' configuration~\cite{2010CQGra..27a5003H}, proposes the installation of two interferometers within each V-shaped arm: one optimized for high-frequency sensitivity and the other for low-frequency sensitivity. Following~\cite{2012PhRvD..86l2001R}, we place the detector at the Virgo site in Cascina, with the first arm of E1 coinciding with the first arm of the Virgo detector. The geometric parameters of the ET interferometers are summarized in Table~\ref{tab:et_constants}, based on the definitions and coordinate data in \cite{T1400308}. The ET is designed to be located underground to mitigate seismic noise. In future rounds, we plan to explore additional configurations, such as two separated L-shaped detectors with different arm lengths.

One of the key advantages of the triangular layout is its ability to produce a \emph{null stream}, which is a linear combination of the three detector outputs that, under ideal conditions, cancel out the GW signal entirely. This cancellation follows from the symmetry of the configuration and the long-wavelength approximation, in which the GW wavelength is much larger than the detector arms. In this regime, the antenna pattern functions of the three detectors combine such that the net GW response vanishes in the null stream. A more detailed discussion of this property will be presented in ~\ref{sec:null-stream}.

\begin{table*}
\centering
\caption{Key geometric and positional parameters of the three 10 km interferometric detectors (E1, E2, E3) that form the Einstein Telescope (ET). All angular values are given in radians. Coordinates are in the Earth-Centered Earth-Fixed (ECEF) frame.}
\label{tab:et_constants}
\begin{tabular}{@{}lccc@{}}
\hline
\textbf{Detectors} & \textbf{E1} & \textbf{E2} & \textbf{E3} \\
\hline
Longitude (rad) & 0.18334 & 0.18406 & 0.18193 \\
Latitude (rad) & 0.76151 & 0.76299 & 0.76270 \\
Elevation (m) & 51.884 & 59.735 & 59.727 \\
X Arm Azimuth (rad) & 0.33916 & 4.52845 & 2.43259 \\
Y Arm Azimuth (rad) & 5.57515 & 3.48125 & 1.38539 \\
X Arm Altitude (rad) & 0.00000 & -0.00078 & -0.00157 \\
Y Arm Altitude (rad) & 0.00000 & -0.00157 & -0.00078 \\
X Arm Midpoint (m) & 5000.0 & 5000.0 & 5000.0 \\
Y Arm Midpoint (m) & 5000.0 & 5000.0 & 5000.0 \\
Vertex Location (X, m) & $4.5464 \times 10^6$ & $4.5394 \times 10^6$ & $4.5424 \times 10^6$ \\
Vertex Location (Y, m) & $8.4299 \times 10^5$ & $8.4507 \times 10^5$ & $8.3564 \times 10^5$ \\
Vertex Location (Z, m) & $4.3786 \times 10^6$ & $4.3854 \times 10^6$ & $4.3841 \times 10^6$ \\
\hline
\end{tabular}
\end{table*}

\subsection{Simulation of the GW Signal for Compact Binary Coalescences}

The simulated signal is composed of gravitational waves (GWs) originating from a cosmological population of compact binary coalescences (CBCs). The intrinsic parameters of these systems are drawn from population-synthesis catalogs produced with the\textbf{ MOBSE} code, as described in \cite{2021MNRAS.505..339M,2022MNRAS.511.5797M} for binary black holes (BBHs) and in \cite{2021MNRAS.502.4877S} for binary neutron stars (BNSs) and black hole–neutron star (BHNS) systems. These catalogs are obtained from the evolution of massive binary stars with primary masses drawn from a Kroupa initial mass function and metallicities in the range $Z=10^{-4}$–$0.02$. The cosmological evolution of the population is included by adopting a metallicity-dependent star-formation rate; for isolated binaries, this follows the Madau \& Fragos model \cite{Madau2017}, with time delays between formation and merger already incorporated in the catalogs.

Following \cite{2020JCAP...03..050M}, we adopt the model with a common-envelope ejection efficiency of 3 for both the BNS and BHNS populations, as this choice reproduces the local merger-rate constraints inferred from the third LIGO-Virgo-KAGRA Gravitational Wave Transcient Catalog (LVK GWTC-3). In this model, neutron-star masses are randomly drawn from a uniform distribution in the range $1.1$–$2.5,M_\odot$, consistent with LIGO–Virgo-KAGRA observations.

For the BBH population, we also follow \cite{2020JCAP...03..050M} and consider a model with a metallicity spread of $\sigma_Z = 0.3$, spin magnitudes drawn from a truncated Maxwellian with $\sigma = 0.1$, and a common-envelope ejection efficiency of 1.0. We further assume the “rapid” supernova model of Fryer et al. (2012, ApJ, 749, 91). The BBH population includes both isolated systems from {\sc MOBSE} and dynamically formed systems, and is able to reproduce the BBH merger rate, as well as the mass and spin distributions constrained by LIGO–Virgo observations (GWTC-3).

\begin{figure}
    \centering
    \includegraphics[width=0.45\textwidth]{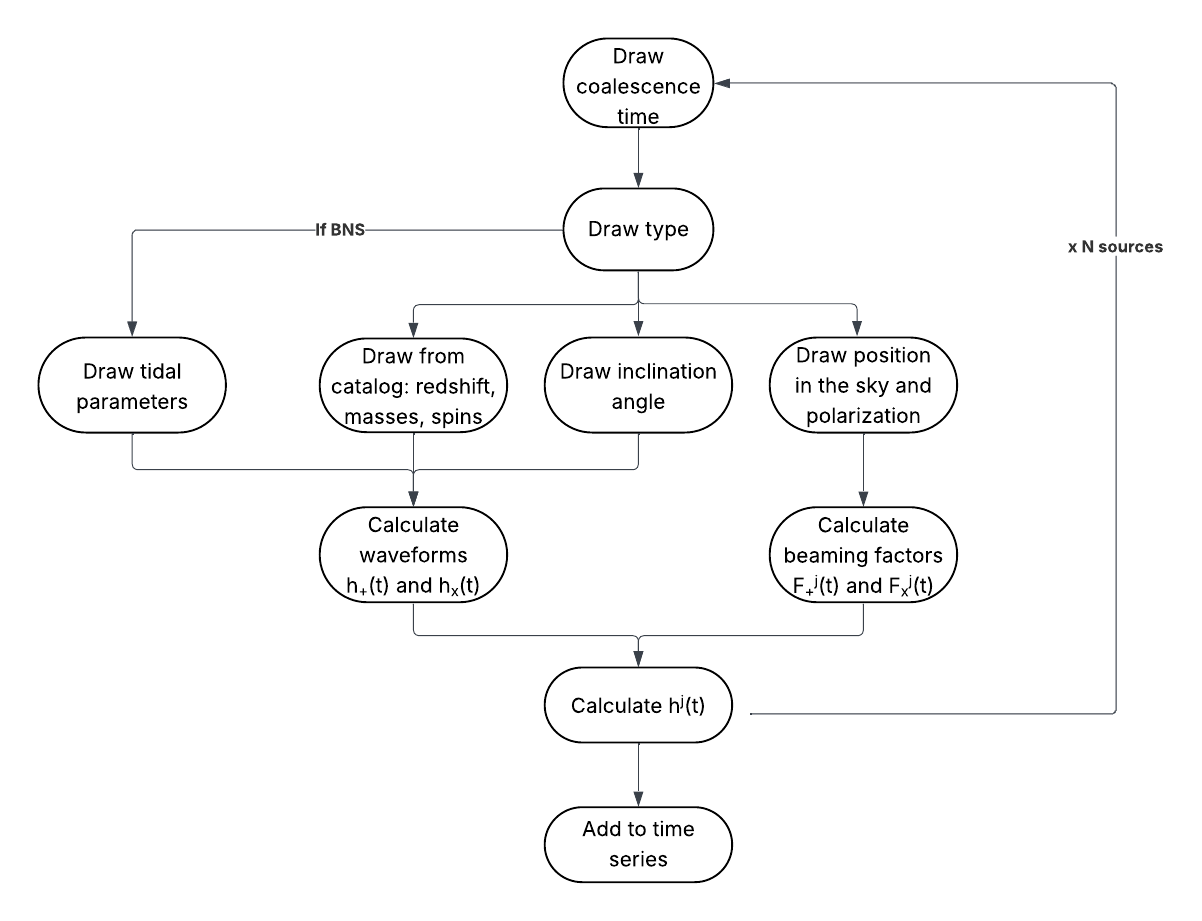}
    \caption{Flowchart summarizing the Monte Carlo procedure used to generate the population of compact binary coalescences and simulate the corresponding ET detector time series.}
    \label{fig-flowchart}
\end{figure}

To generate a population of compact binaries, we employed Monte Carlo techniques.
The full procedure is summarized in the flowchart shown in Fig.~\ref{fig-flowchart} and is detailed step by step below.
For each source, the following steps were implemented:

\begin{itemize} 

\item The time interval between successive coalescences was modeled as a Poisson process, with the time $\tau$ drawn from an exponential distribution $P(\tau) = \exp(- \tau / \lambda)$, where $\lambda = 38$ s is the average time between events. This value was directly calculated from the catalog as the ratio of the catalog's duration to the number of events.

\item The binary type was selected with the following distribution: 87\% BNSs, 3\% BHNSs, and 10\% BBHs. These proportions were taken directly from the catalogs, yielding average waiting times of 43.7 s for BNSs, 390.8 s for BBHs, and 1371.2 s for BHNSs.

\item The masses, spins, distances, and redshifts for each source were drawn from the corresponding catalog. The luminosity distance was computed from the redshift using a flat $\Lambda$-Cold Dark Matter (CDM) cosmology, based on the Planck 2015 results, with $\Omega_M = 0.308$ for the matter density and $H_0 = 67.8$ km s$^{-1}$Mpc$^{-1}$ for the Hubble constant \cite{2016A&A...594A..13P}.

\item The inclination angle $\iota$ was drawn from an isotropic distribution, while the polarization angle $\psi$ and the initial phase $\phi_0$, when entering the frequency band at 5 Hz, were drawn from uniform distributions.

\item The right ascension $\alpha$ was drawn from a uniform distribution in $[0, 2\pi)$, and the declination $\delta$ was drawn such that $\cos(\delta)$ is uniform in $[-1, 1]$, ensuring an isotropic distribution over the celestial sphere.

\item 
For BNS systems, the tidal deformability parameters were computed using the DD-LZ1 equation of state (EOS) \footnote{See \url{https://compose.obspm.fr/eos/280} for details}. 
DD-LZ1 is a density-dependent relativistic mean-field EOS that produces neutron-star radii and maximum masses consistent with current astrophysical constraints. 
For each neutron-star mass in the catalog, the tidal Love number $k_2$ and the radius $R$ were obtained directly from the tabulated EOS and the dimensionless tidal deformability was calculated from the expression:
\begin{equation}
\Lambda = \frac{2}{3} k_2 \left( \frac{R c^2}{G M} \right)^5 .
\end{equation}
Over the mass range used here ($1.1$--$2.5\,M_\odot$), DD-LZ1 yields tidal deformability $\Lambda \sim 10-4000$, decreasing with mass.

\item The two polarizations $\tilde{h}_+(\Theta, f)$ and $\tilde{h}_{\times}(\Theta, f)$ were generated using the IMRPhenomPv2 waveform model with tidal effects (NRTidalv2) for BNSs, and the IMRPhenomXPHM model for both BBHs and BHNSs, including higher modes. These frequency-domain waveforms were then transformed into time-domain signals $h_+(\Theta, t)$ and $h_{\times}(\Theta, t)$ using fast Fourier transforms. The set of parameters $\Theta$ includes the component masses, the spins of the BHs for BBHs and BHNSs, the tidal parameters for the NSs in BNS systems, the inclination angle, the initial phase, and the luminosity distance. The masses used are ``redshifted" or ``observed" masses, corrected by a factor of $(1+z)$.

\item The time-dependent antenna pattern functions for the three ET
detectors are denoted by
$F_+^j(\alpha, \delta, \psi; t)$ and
$F_\times^j(\alpha, \delta, \psi; t)$, where the superscript
$j = 1,2,3$ labels the detector (E1, E2, and E3), and
$(\alpha, \delta, \psi)$ are the source right ascension, declination,
and polarization angle.
These functions were calculated based on the detector positions and
orientations, and describe the response of detector $j$ to the two GW
polarizations.

Fig.~\ref{fig-responses} illustrates the antenna pattern functions
$F_+(t)$ and $F_\times(t)$ for the three ET interferometers as a function
of sidereal time, for a random sky location. Due to the Earth's
rotation, the sensitivity of each detector to a given source direction
varies over time, producing the characteristic modulation shown in the
plots.

\begin{figure}
    \centering
    \includegraphics[width=0.45\textwidth]{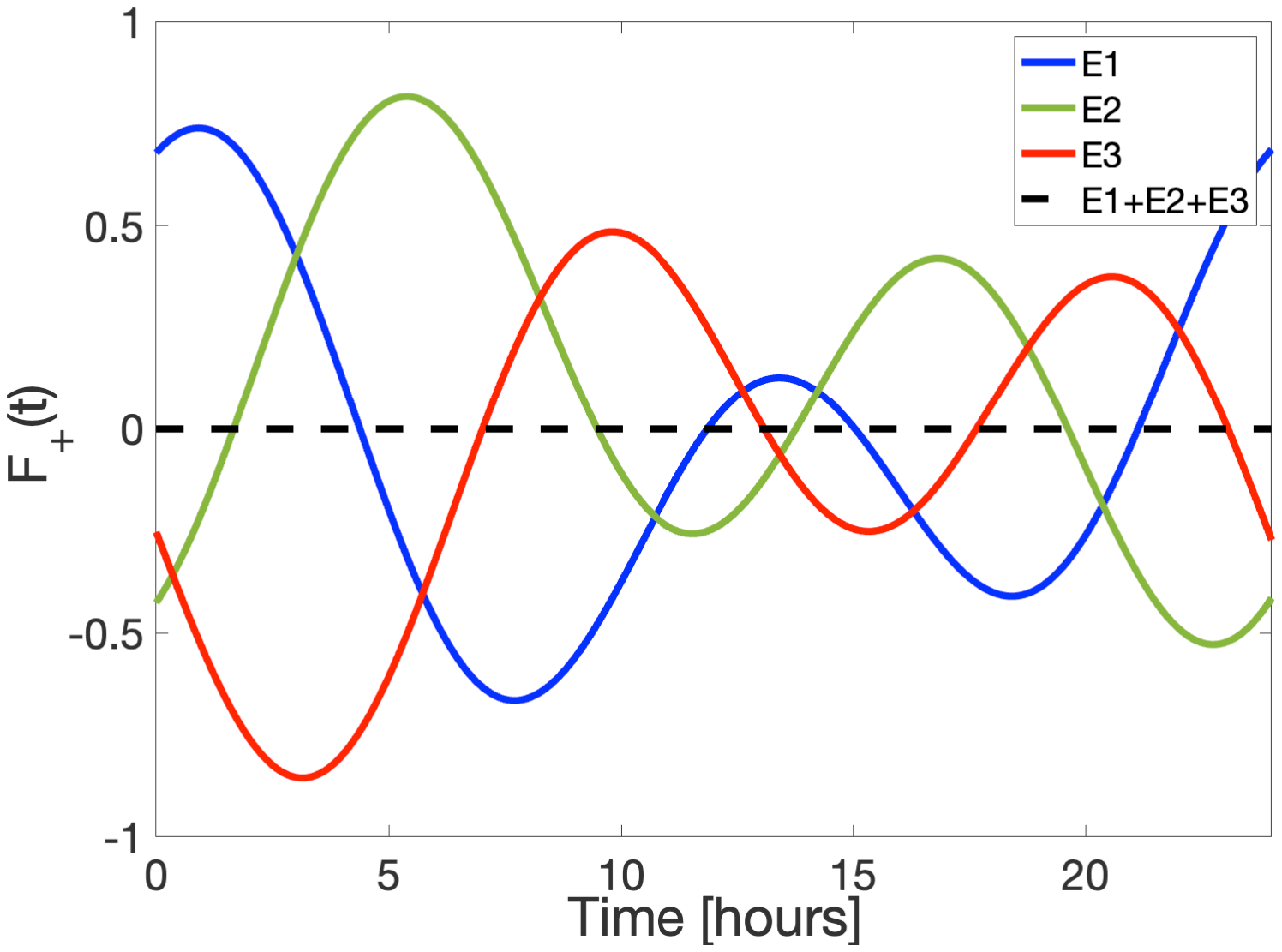}
    \includegraphics[width=0.45\textwidth]{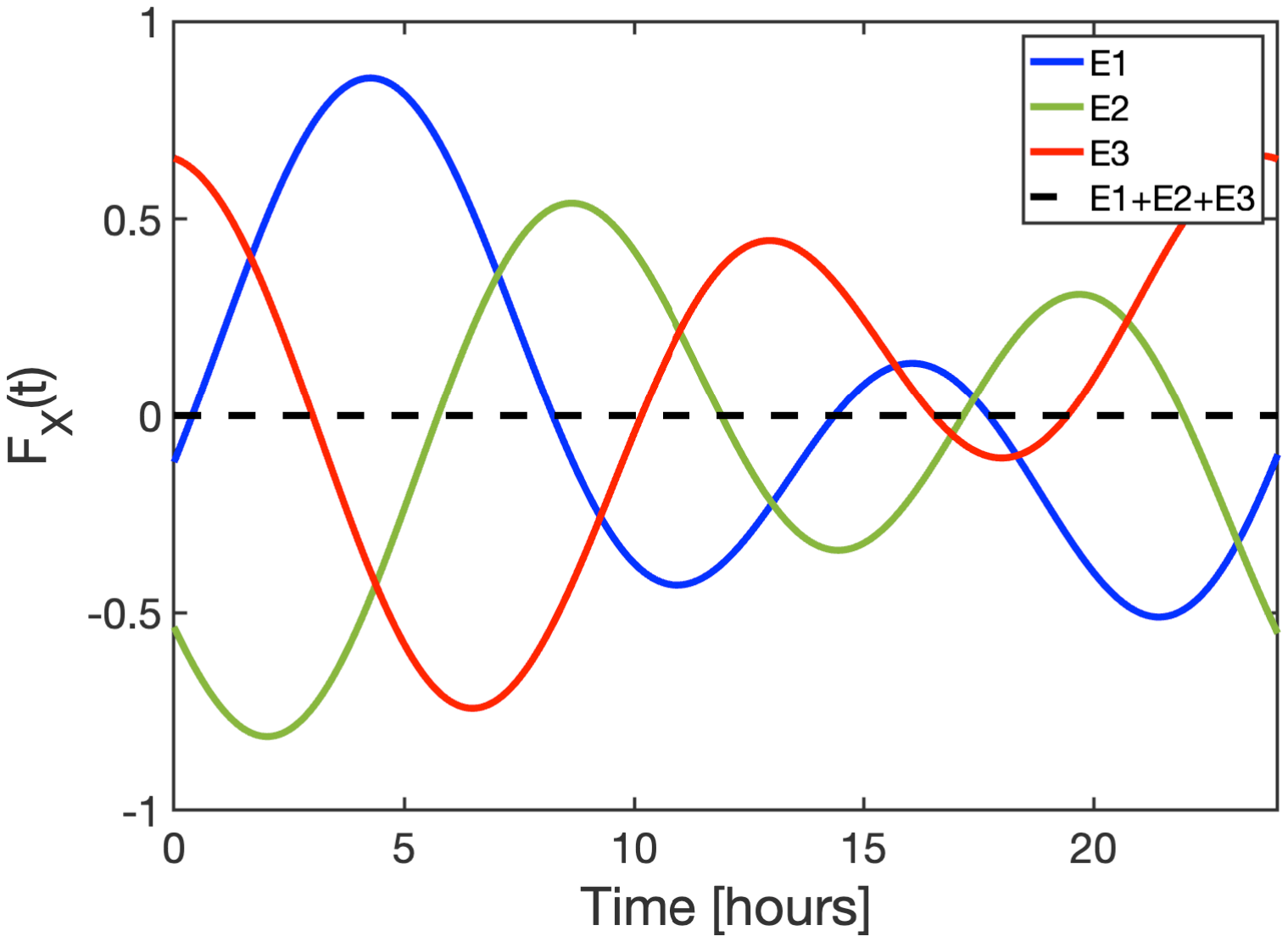}\\
    \caption{Antenna pattern functions \( F_+(t) \) (top) and \( F_\times(t) \) (bottom) over one sidereal day for the three ET interferometers and the sum E1+E2+E3.}
    \label{fig-responses}
\end{figure}

\item The responses $h^j(t)$ were then computed as 
\begin{equation}
h^j(t) = F^j_+(t) \, h_+\big(t - t_d^j\big) + F^j_{\times}(t) \, h_{\times}\big(t - t_d^j\big),
\end{equation}
where $t_d^j$ represents the wave travel time between the detector and the Earth's center. 
The signals from all sources were then added, denoted by the summation over $k$, to form the final GW time series:
\begin{equation}
h^j_{\mathrm{GW}}(t) = \sum_k h^j_k(t).
\label{eq:hGW}
\end{equation}

Note that the simulated GW signals evolve continuously across successive data segments, meaning that when a source begins in one segment, it is seamlessly continued into the next.

Figure~\ref{fig-GWsignal} shows the GW signal for the first 2048-second data segment in E1. The signal consists of an intermittent component from BBH mergers and occasional neutron star–black hole systems, superimposed on a continuous background signal from lower-amplitude BNS mergers (shown in blue).

\begin{figure} 
\centering 
\includegraphics[width=0.5\textwidth]{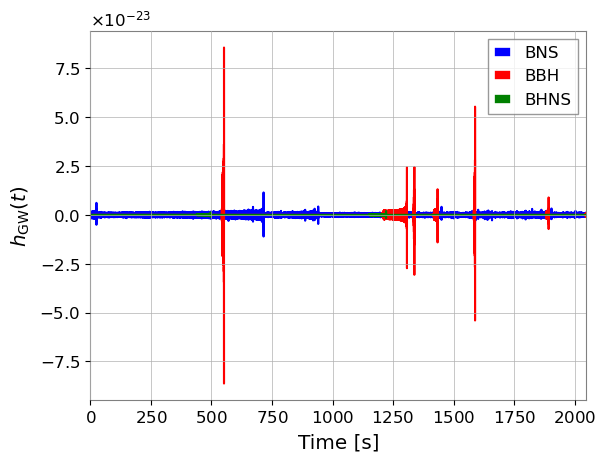} 
\caption{Time series of the GW signal within the first 2048-second data segment sampled at 8192 Hz. The signal from BNS is shown in blue, BHNS in green, and BBH in red.} 
\label{fig-GWsignal}
\end{figure}

 \end{itemize}
 
\subsection{Simulation of the noise}
\begin{figure} 
\centering 
\includegraphics[width=0.5\textwidth]{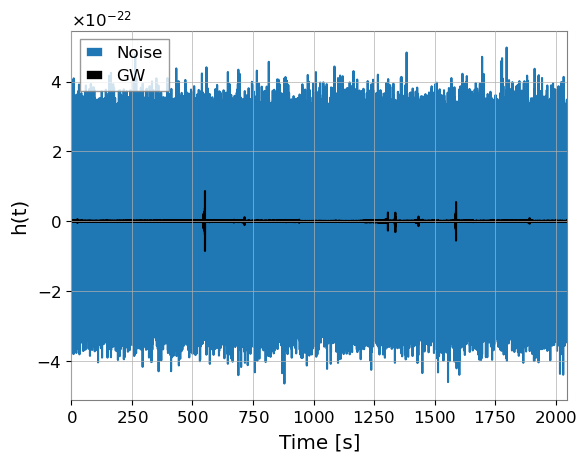} 
\caption{Time series of the first 2048-second data segment sampled at 8192 Hz. The blue curve shows a realization of detector noise, while the black curve represents the GW signal from CBCs, as also shown in Fig.~\ref{fig-GWsignal}.} 
\label{fig-signal} 
\end{figure}

Assuming no correlated instrumental or environmental noise, independent Gaussian frequency series with zero mean and unit variance were generated for each of the three ET detectors. 
This simplified noise model is intentionally adopted for \textit{ET-MDC1} to establish a controlled baseline. In realistic conditions, correlated noise between co-located detectors and non-Gaussian transients are expected to impact pipeline performance, for instance by reducing the effectiveness of null-stream consistency tests and increasing false-alarm rates. A quantitative assessment of these effects is deferred to future mock data challenges.
The frequency series were then colored using the noise power spectral density (PSD) and inverse Fourier transformed to the time domain.
To ensure continuity between consecutive segments, the time-domain data were combined using overlapping and blending techniques \cite{lalsuite}. Specifically, each segment overlapped by 50\%, with half of the data from one segment overlapping with the next. In the overlapping regions, the old segment’s samples were multiplied by a cosine function (decreasing from 1 to 0), and the new segment’s samples by a sine function (increasing from 0 to 1), evaluated such that the squares of the two weights sum to one at each point. This energy-preserving blend ensures smooth transitions between segments and avoids discontinuities at the boundaries.
Fig.~\ref{fig-signal} shows the time series of the first data segment in E1, where the GW signal (in black), the same shown in Fig.~\ref{fig-GWsignal}, is buried within the simulated detector noise (in blue).

\begin{figure}
\centering
\includegraphics[width=0.5\textwidth]{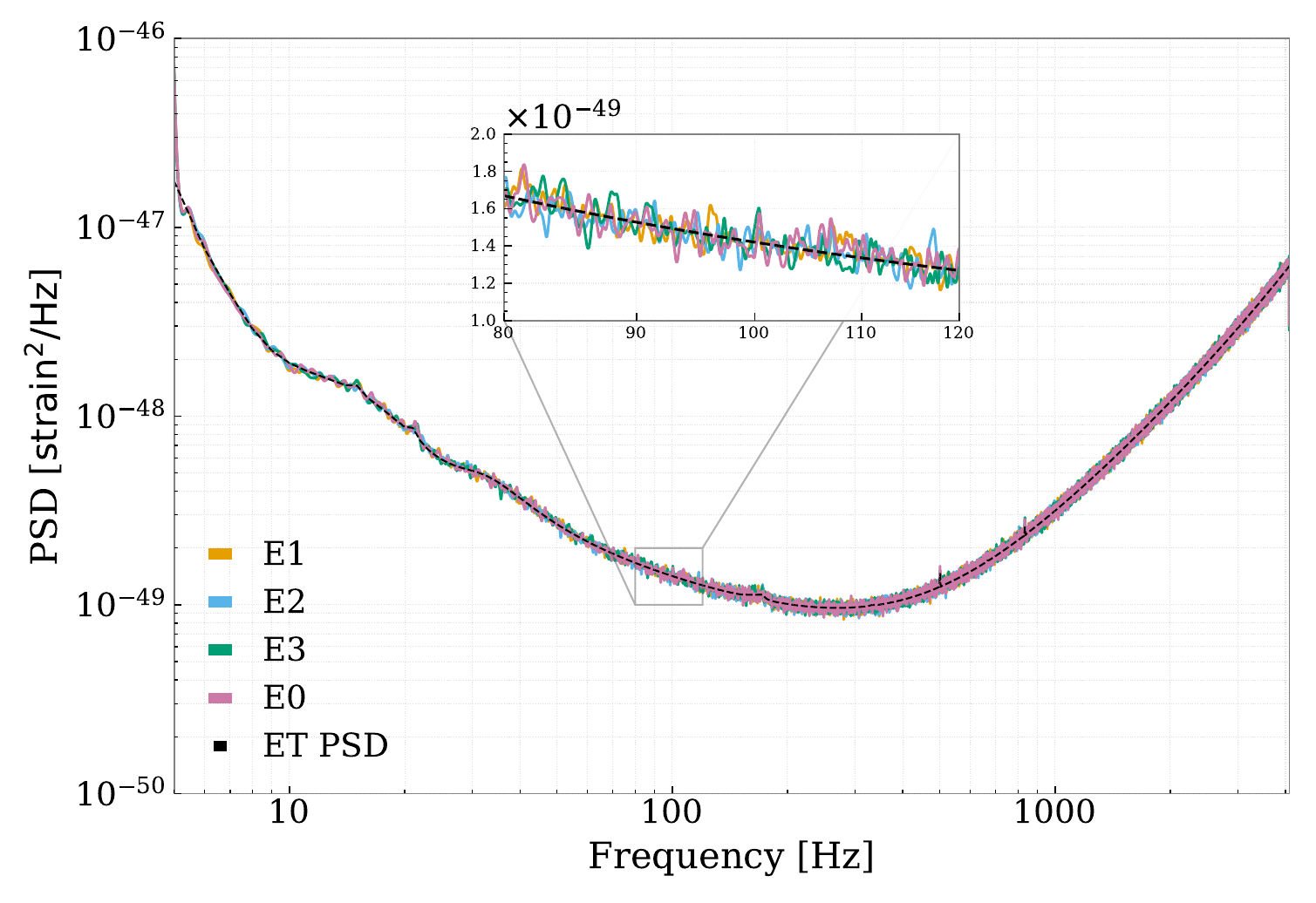}
\caption{Comparison between the predicted noise power spectral density (dashed black line) and the PSD estimated from the first 2048-second segment of simulated ET data for E1, E2, E3, and the null stream E0. The agreement confirms the fidelity of the noise simulation process and indicates that the null stream cancels the GW signal while retaining the instrumental noise characteristics.}
\label{fig-psdcomp}
\end{figure}

To validate the accuracy of the noise simulation, we estimated the power spectral density (PSD) from the first 2048-second segment of the simulated dataset, using Welch’s method with 4-second Hann-windowed, overlapping segments. This approach balances frequency resolution with reduced variance by averaging over many short segments. 
As shown in Fig.~\ref{fig-psdcomp}, the recovered PSD for the three detectors E1, E2, and E3 closely follows the predicted sensitivity curve across the relevant frequency range.

Although the underground placement of ET is expected to reduce seismic noise compared to surface detectors like LIGO, Virgo, or Kagra, the assumption of uncorrelated may be optimistic due to the near co-location of the three ET detectors. The next rounds of ET mock data challenges will incorporate a more detailed study of environmental noise effects \cite{2022PhRvD.106d2008J,2023EPJP..138..352J,2023arXiv230502694J,2024PhRvD.109j2002J}.

\subsection{Null Stream}
\label{sec:null-stream}
In the simulation, we also compute the null stream, designed to cancel out
the GW signal, ideally leaving behind only noise.

The null stream is formed by combining the outputs of the three ET detectors
(E1, E2 and E3) as
\begin{equation}
s_0(t) = \frac{s^1(t) + s^2(t) + s^3(t)}{\sqrt{3}} .
\end{equation}

Each detector output $s^j(t)$ contains the superposition of GW signals from
multiple sources, as well as the instrumental noise:
\begin{equation}
s^j(t) = h_\mathrm{GW}^{j}(t) + n^j(t),
\end{equation}
where $h_\mathrm{GW}^{j}(t)$ is given in Eq.~\ref{eq:hGW}.

In the long-wavelength approximation, the ET triangular geometry implies
(see \cite{2012PhRvD..86l2001R,2022PhRvD.105l2007G,2023EPJP..138..352J} for details)
\begin{equation}
F_A^1(\alpha_k, \delta_k,\psi_k;t)
+ F_A^2(\alpha_k, \delta_k,\psi_k;t)
+ F_A^3(\alpha_k, \delta_k,\psi_k;t) = 0,
\end{equation}
where $A=+,\times$, so that the GW contribution from each individual source
cancels, leaving only the noise:
\begin{equation}
s_0(t) \simeq
\frac{n^1(t) + n^2(t) + n^3(t)}{\sqrt{3}} .
\end{equation}

\begin{figure}
\centering
\includegraphics[width=\columnwidth]{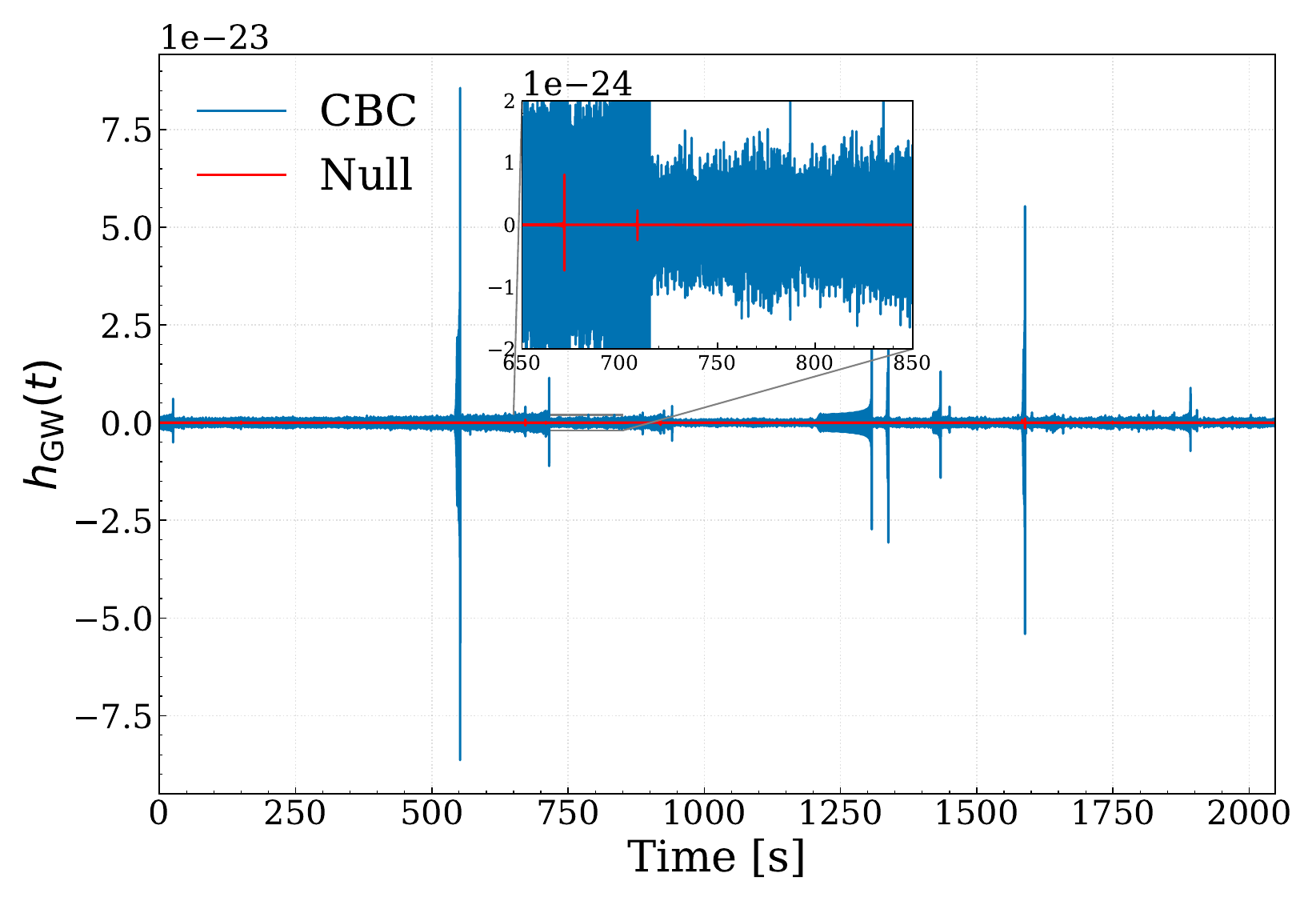}
\caption{Time series of the GW signal (in blue) within the first 2048-second data segment sampled at 8192 Hz. The null stream is indicated in red.}
\label{fig-nscbc}
\end{figure}

A small residual may remain due to the breakdown of the long-wavelength approximation and slight differences in the detectors’ PSDs, particularly for high-frequency or very loud signals. As shown in Figure~\ref{fig-nscbc}, this residual is minor and does not significantly affect the noise properties of the null stream.

Figure~\ref{fig-nscbc} shows the null stream computed for a frame containing only a coherent GW signal (i.e., without injected noise), illustrating significant suppression of the signal. However, a small residual remains visible.

This residual is a known effect that arises from the breakdown of the long-wavelength approximation, which assumes that the GW wavelength is much larger than the detector’s dimensions. In practice, the detectors are located at the vertices of an equilateral triangle with 10 km sides, leading to small time delays between the detectors as the GW passes through the array. These delays result in imperfect cancellation of the GW signal in the null stream, particularly for high-frequency components or very loud signals \cite{2022PhRvD.105l2007G}. Other factors, such as differences in the detectors’ noise power spectral densities, can also impact the null stream’s performance \cite{2023EPJP..138..352J}. Despite this limitation, the power spectral density of the null stream, shown in Figure~\ref{fig-psdcomp}, closely follows the ET sensitivity curve, confirming that the GW signal is effectively suppressed while the noise characteristics of the detectors are preserved.

\section{Statistics of the simulated population}

This section presents the statistical properties of the GW sources in ET-MDC1, including population distributions and detection-related metrics. The goal is to provide a quantitative overview of the parameter space covered in our dataset. For a detailed discussion of the astrophysical models underlying these simulations, we refer readers to \cite{2021MNRAS.505..339M,2022MNRAS.511.5797M,2021MNRAS.502.4877S}.

\subsection{Population properties}

\begin{itemize}[label={}]
\item \textbf{Component masses:}
Fig.\ref{fig-mass} shows the distribution of primary mass ($m_1$) versus secondary mass ($m_2$) for the simulated BNSs, BBHs, and BHNSs. This visualization provides insight into the range of masses covered in the dataset and serves as a reference for matched-filtering template coverage.

\begin{figure}
    \centering
    \includegraphics[width=0.5\textwidth]{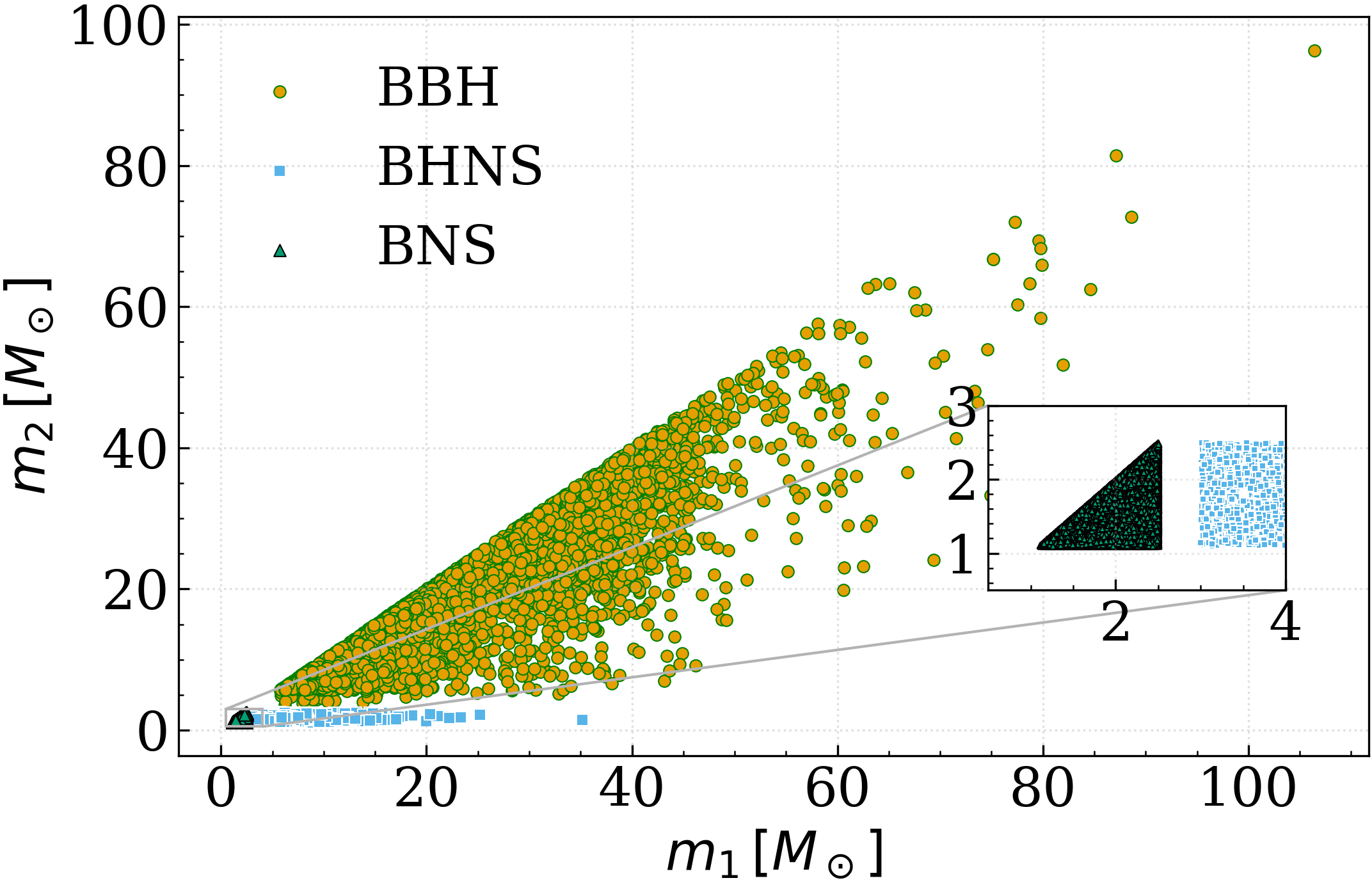}
    \caption{Scatter plot of the primary mass ($m_1$) versus the secondary mass ($m_2$) for the sources in our dataset. The 61031 BNSs are indicated in green, the 2025 BHNSs in blue, and the 6725 BBHs in yellow.}
    \label{fig-mass}
\end{figure}

\item \textbf{Component spins:}
Fig.\ref{fig-spin} presents the distribution of dimensionless spin magnitudes, $\chi_1 = s_1/m_1$ versus $\chi_2 = s_2/m_2$, for BBH systems. The majority of sources cluster in the low-spin regime ($\chi_1, \chi_2 \leq 0.4$). Additionally, a subset of systems features higher primary spins ($\chi_1 \sim 0.7 - 0.75$), with secondary spins either in the same range or lower ($\chi_2 \leq 0.4$). These distributions reflect the properties of the population synthesis models used in the simulation.

\begin{figure}
    \centering
    \includegraphics[width=0.45\textwidth]{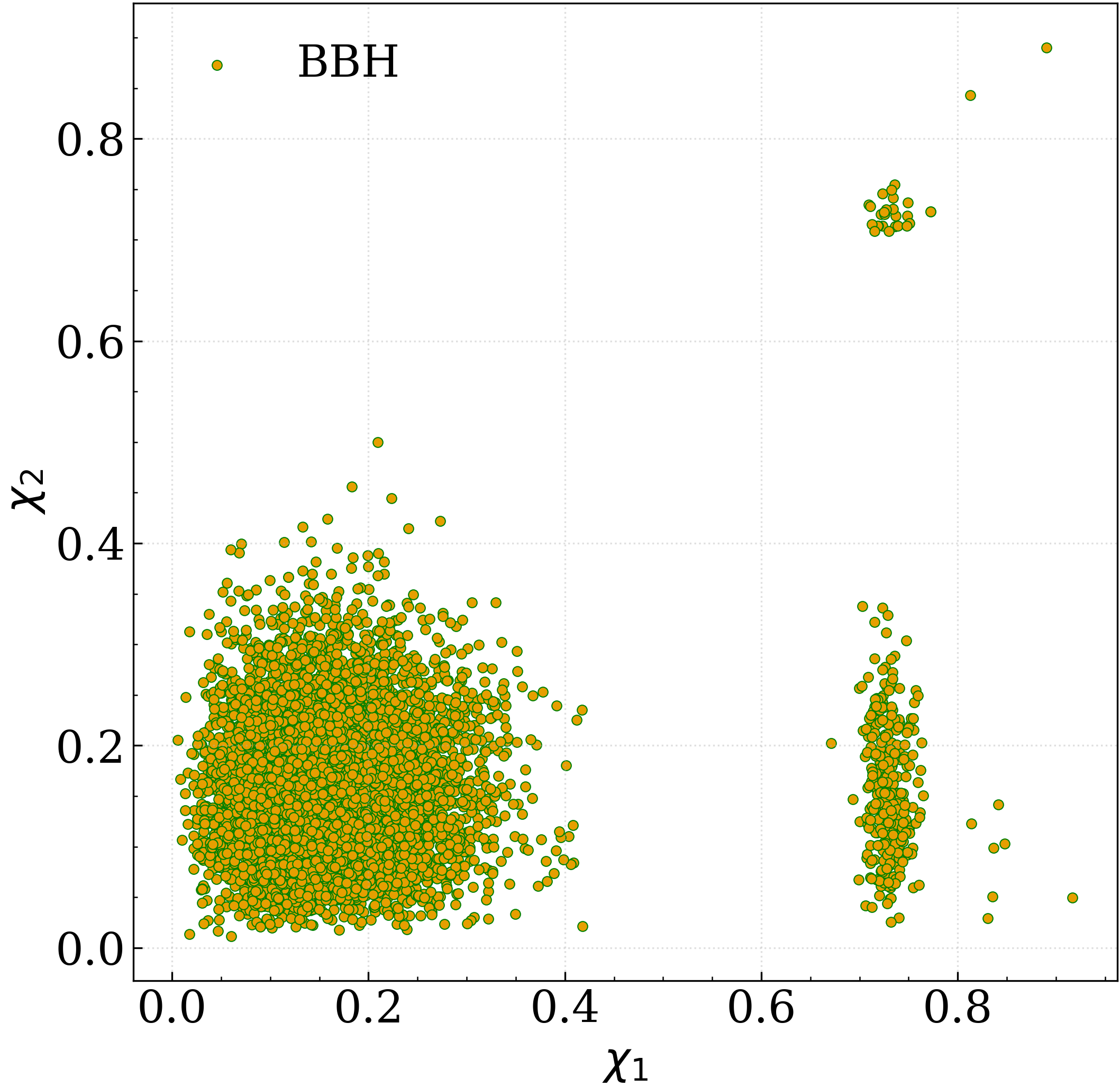}
    \caption{Scatter plot of the dimensionless spin magnitudes, $\chi_1 = s_1/m_1$ versus $\chi_2 = s_2/m_2$ for the 6725 BBHs in our dataset.}
    \label{fig-spin}
\end{figure}

\item \textbf{Tidal deformability:} For BNS systems, the tidal deformability parameter $\Lambda$ depends on the neutron star equation of state (EoS) and impacts the waveform through finite-size effects. Fig.\ref{fig-tidal} shows the tidal deformability parameter as a function of mass for the BNS population. The distribution is consistent with the chosen EoS (DD-LZ1) used in the simulation.

\begin{figure}
    \centering
    \includegraphics[width=0.5\textwidth]{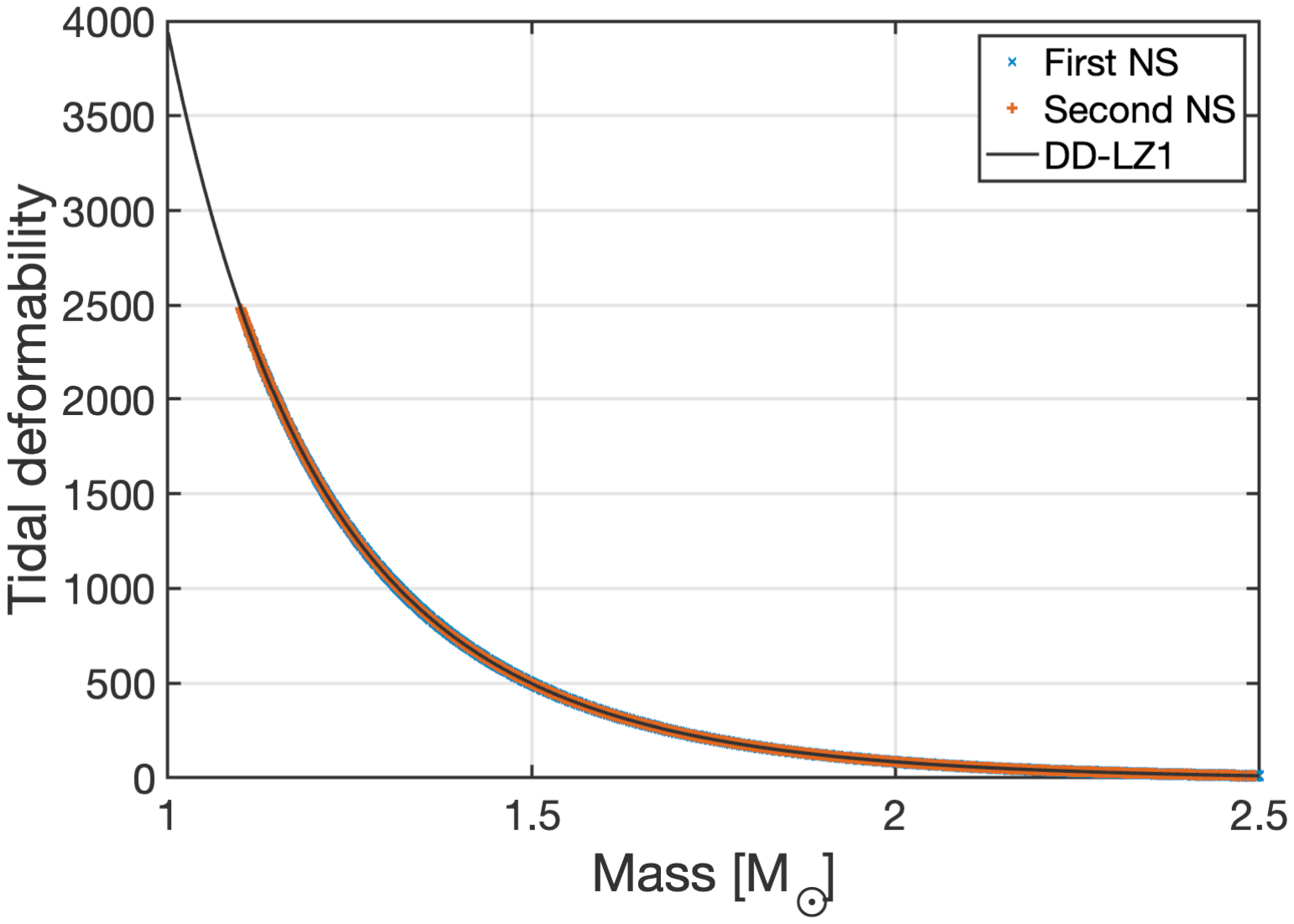}
    \caption{Tidal deformability parameter $\Lambda$ as a function of neutron-star mass for the binary neutron star population in the synthetic catalog. The blue and red crosses correspond to the individual component masses of the first and second neutron stars respectively. The solid black curve shows the $\Lambda(M)$ relation predicted by the DD--LZ1 equation of state used in the simulations.}
    \label{fig-tidal}
\end{figure}

\item \textbf{Redshift:}
Fig.\ref{fig-redshift} presents the redshift distribution of the simulated sources. The population peaks around $z \sim 2$, consistent with the maximum of the cosmic star formation rate.

\begin{figure}
    \centering
    \includegraphics[width=0.5\textwidth]{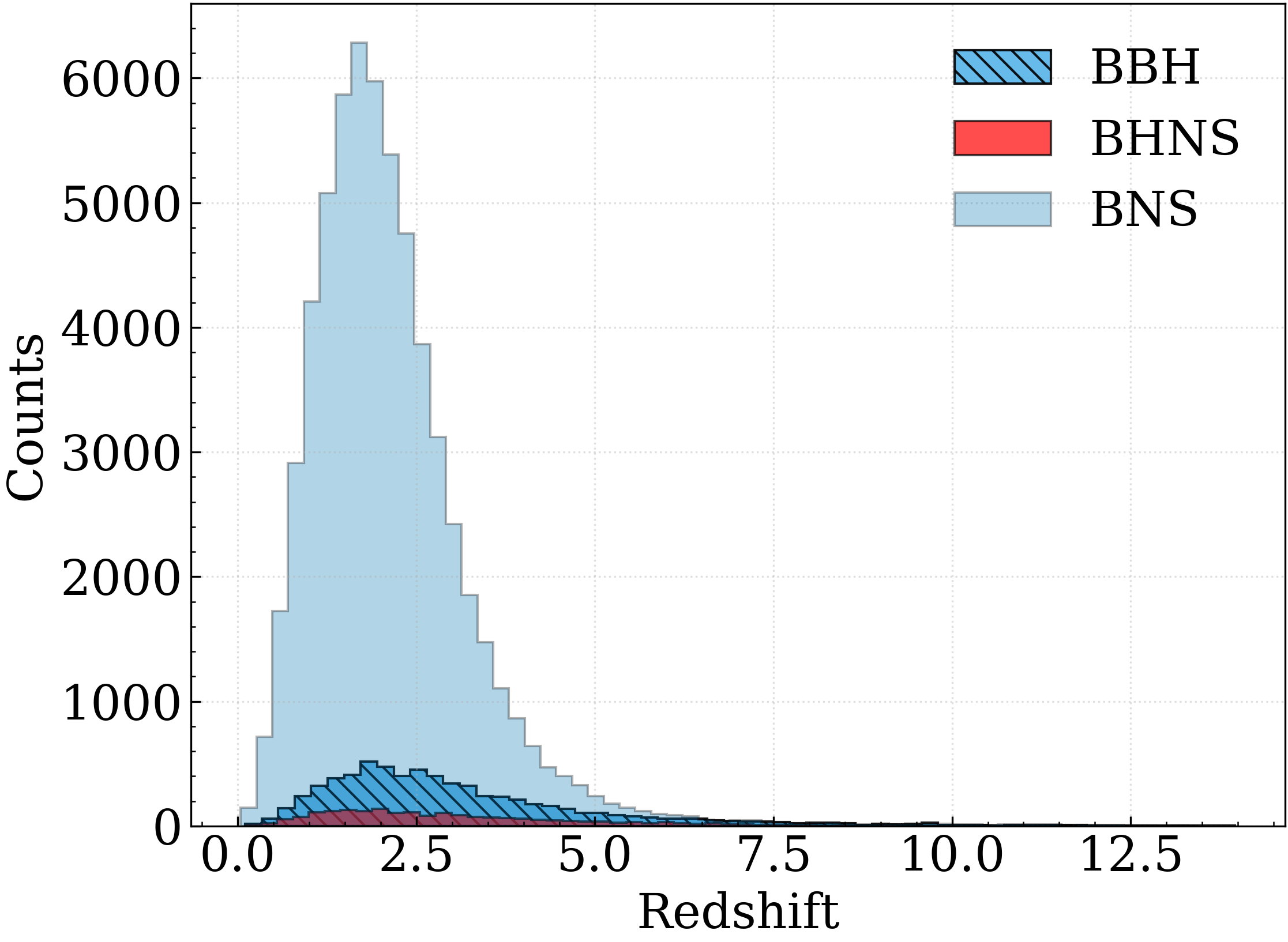}
    \caption{Histogram of redshifts for the sources in our dataset. BNSs are shown in blue, BHNSs in green, and BBHs in red.}
    \label{fig-redshift}
\end{figure}


\item \textbf{Signal duration:}
The inspiral duration from a reference frequency $f_\mathrm{low}$ to the merger, ignoring the spin, is often approximated as
\begin{equation}
t_\mathrm{insp} \approx \frac{5c^5}{256\pi^{8/3}G^{5/3}} \mathcal{M}^{-5/3}f_\mathrm{low}^{-8/3} ,
\end{equation}
where $\mathcal{M}$ is the redshifted chirp mass, defined as $\mathcal{M} = (1 + z)\, \frac{(m_1 m_2)^{3/5}}{(m_1 + m_2)^{1/5}}$.
Lower-mass systems, such as BNSs, produce long-duration signals that can last up to nearly two hours, while high-mass BBHs generate much shorter signals. For instance, a BNS with an intrinsic total mass of $M=2.7\,M_\odot$ at $z=0.2$ has a duration of nearly 2 hours, whereas a BBH with $M=70\,M_\odot$ at $z=11.7$ produces a signal lasting only about 1.8 seconds.

In Fig.~\ref{fig-duration}, we show the distribution of signal durations in our dataset, calculated directly from the full waveforms and including the merger and ringdown phases. 

\begin{figure}
    \centering
    \includegraphics[width=0.5\textwidth]{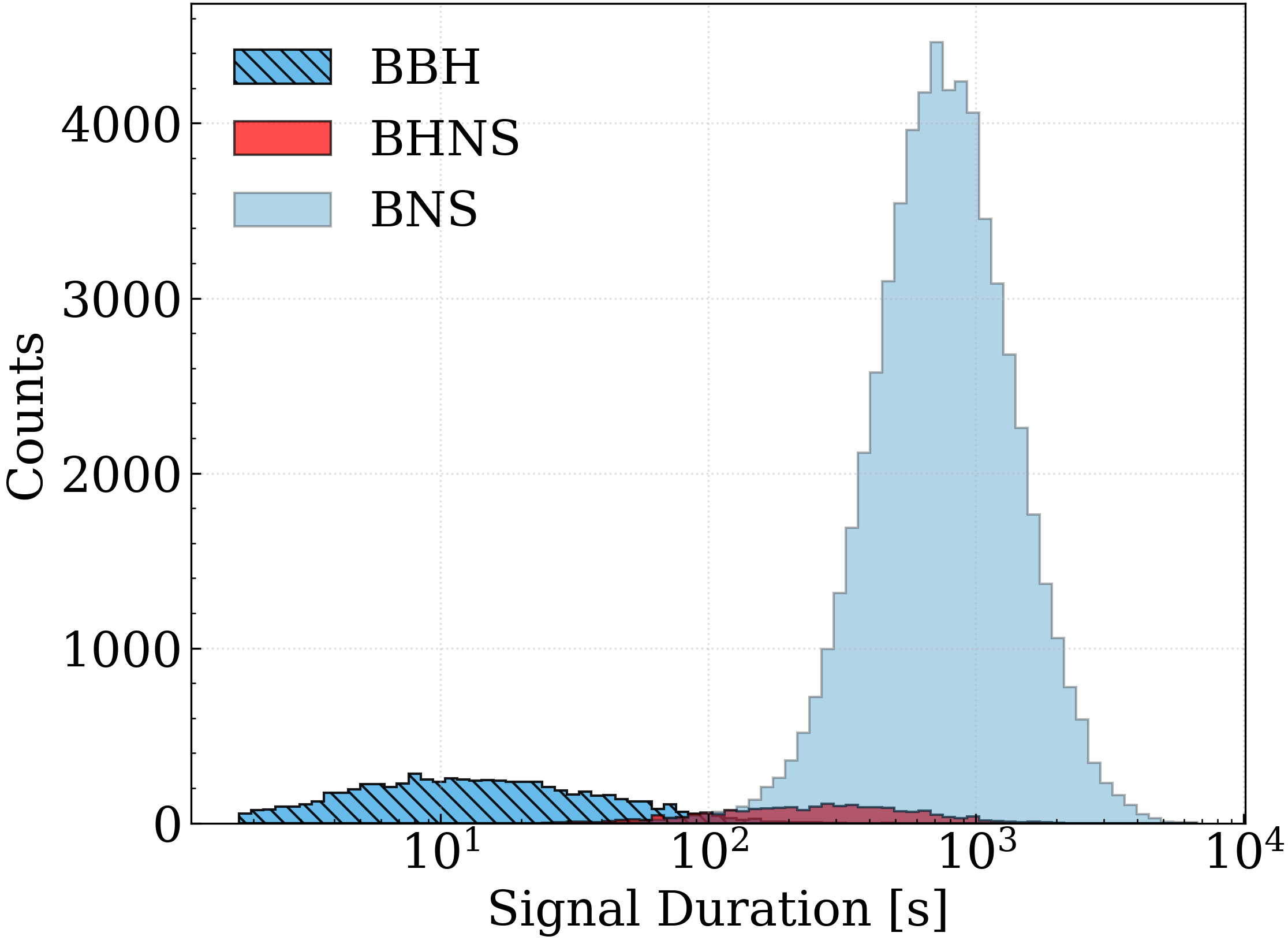}
    \caption{Histogram of duration (in seconds), in logarithmic scale for the sources in our dataset. BNSs are shown in blue, BHNSs in green, and BBHs in red.}
    \label{fig-duration}
\end{figure}

\item \textbf{Duty cycle:}
the duty cycle characterizes the overlap between GW signals and is defined as the ratio of the average signal duration to the average time interval between successive events:
\begin{equation}
DC = \frac{t_{\mathrm{wform}}}{\lambda}
\end{equation}
where $t_{\mathrm{wform}}$ is the average waveform duration, and $\lambda$ is the mean interval between events.

Table~\ref{tab:duty_cycle} lists $t_{\mathrm{wform}}$, $\lambda$, and the resulting duty cycle $DC$ for the three source populations: BNS, BHNS, and BBH. The BNS population exhibits a continuous signal with a duty cycle of 20, while BHNS and BBH populations show intermittent signals with duty cycles well below 1.

In Fig.\ref{fig:duty_cycle}, we show the duty cycle as a function of the SNR threshold. From the plot, we observe a turning point at an SNR threshold of 18, where the duty cycle reaches 1, marking the transition from continuous to non-continuous behavior. Below SNR = 18, there is a clear risk of overlap between sources, although the chance of having two mergers in a 1-second window is effectively zero in the dataset.

\begin{table}
    \centering
    \begin{tabular}{|c|c|c|c|c|}
        \hline
        \textbf{Parameter} & \textbf{BNS} & \textbf{BHNS} & \textbf{BBH} & \textbf{Total} \\ \hline
        $t_{\mathrm{wform}}$ (s) & 891 & 334 & 23 & 790 \\ \hline
        $\lambda$ (s)            & 44  & 1371 & 391 & 38 \\ \hline
        $DC$                     & 20  & 0.24 & 0.06 & 21 \\ \hline
    \end{tabular}
    \caption{Comparison of waveform duration $t_{\mathrm{wform}}$, average inter-event time $\lambda$, and duty cycle $DC$ across source types.}
    \label{tab:duty_cycle}
\end{table}

\begin{figure}[h!]
    \centering
    \includegraphics[width=0.5\textwidth]{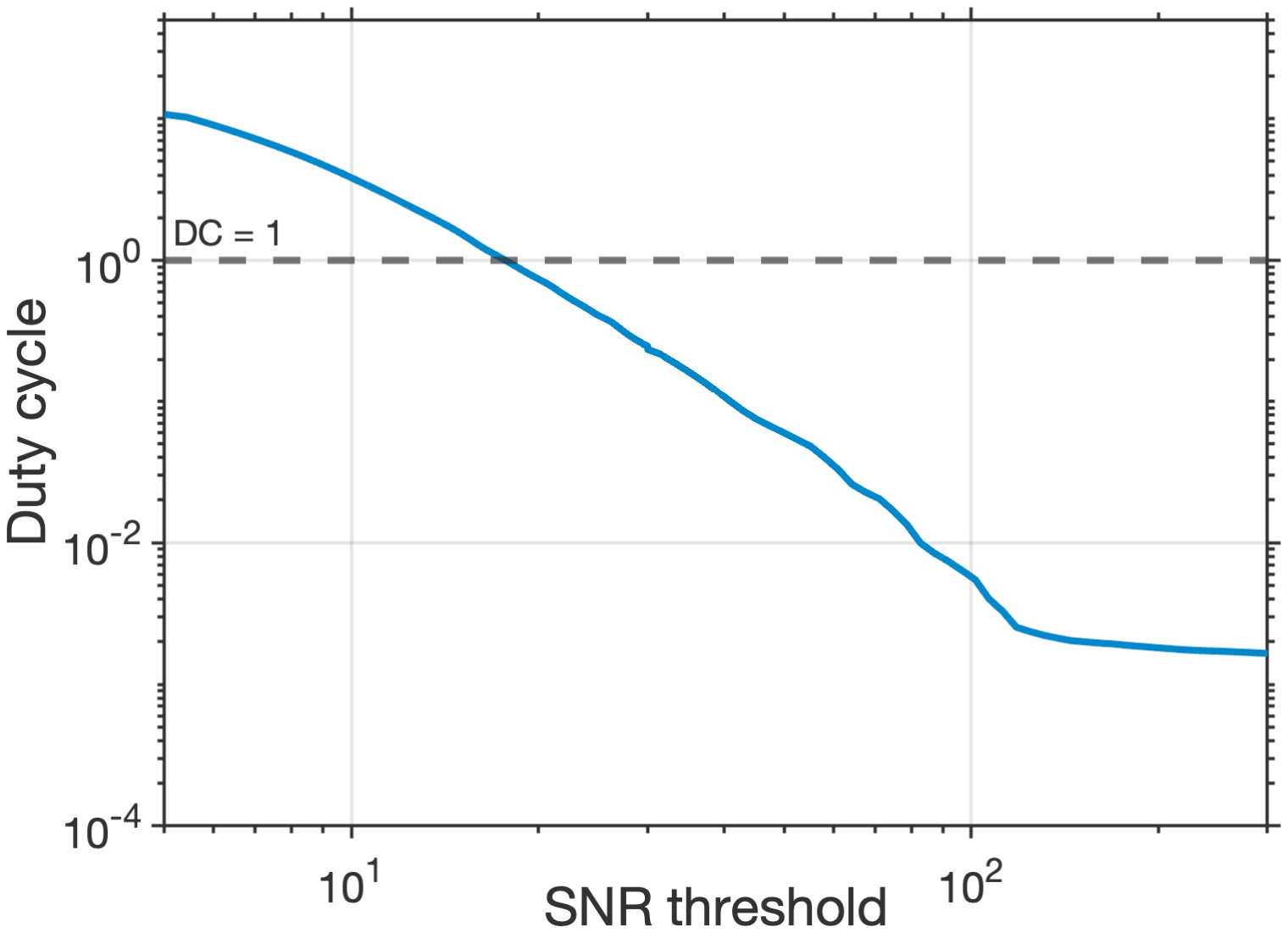}  
    \caption{Duty cycle as a function of the SNR threshold. The plot shows how the duty cycle varies with the SNR threshold. We observe that an SNR threshold of 18 represents the turning point between continuous and non-continuous behavior, with a duty cycle of 1.}
    \label{fig:duty_cycle}
\end{figure}

\end{itemize}

\subsection{Detection and Selection Effects}

With its tenfold improvement in sensitivity over current detectors, ET will probe a volume roughly a thousand times larger, enabling the detection of a significantly greater fraction of compact binary mergers across cosmic time. This increase in detection range will provide valuable insights into the formation and evolution of these systems.

This section examines the ability of the ET to detect GW signals from the simulated population. The efficiency of detection depends primarily on the signal-to-noise ratio (SNR), which quantifies the strength of a GW signal relative to the detector noise. 

For a single detector, the optimal SNR is given by the analytic form:  

\begin{equation}
  \rho_{j}=\sqrt{4 \int_0^\infty \frac{|\tilde{h_j}(f)|^2}{S_n(f)} \, df}
\label{eq-snr} 
\end{equation}  

where $\tilde{h_j}(f)$ is the Fourier transform of the GW signal in detector $j$, and $S_n(f)$ is the one-sided power spectral density of the detector noise.  

For a network of three detectors (E1, E2, and E3), the total SNR is obtained by summing the individual contributions in quadrature:  

\begin{equation}
\text{SNR}_{\text{tot}}^2 = \rho_{1}^2 + \rho_{2}^2 + \rho_{3}^2.
\label{eq-snrtot} 
\end{equation}  

We assume that a source is detected if the combined SNR is above a threshold of 8, while an SNR of 12 is required for precision studies.  

\begin{figure}
    \centering
    \includegraphics[width=0.5\textwidth]{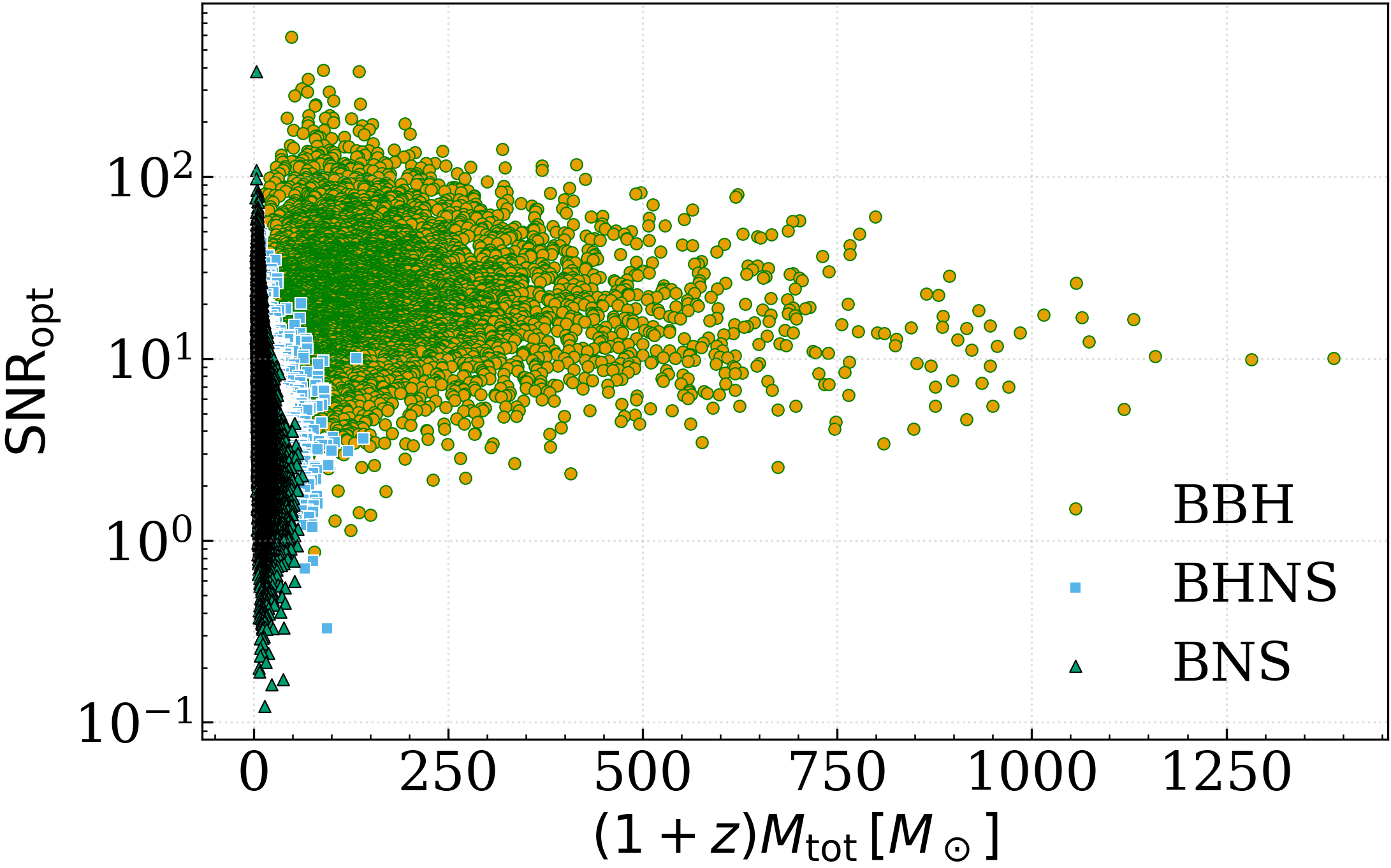}
    \caption{Optimal signal-to-noise ratio ($\rho_{opt}$) as a function of the observed total mass $M_{z,tot}$.}
    \label{fig-snr_opt}
\end{figure}

Fig.~\ref{fig-snr_opt} shows the optimal SNR as a function of the observed total mass, $M_{z,tot} = (m_1 + m_2)(1+z)$. 

The expected fraction of detected sources in our dataset is as follows:  

\begin{itemize}[label=\scriptsize$\bullet$] 
    \item 61031 BNSs: 19\% have $\text{SNR} > 8$, 7\% have $\text{SNR} > 12$.  
    \item 2025 BHNSs: 27\% have $\text{SNR} > 8$, 12\% have $\text{SNR} > 12$.  
    \item 6725 BBHs: 91\% exceed $\text{SNR} = 8$, 79\% exceed $\text{SNR} = 12$, and about 152 BBHs have an SNR above 100, making them ideal for high-precision studies.  
\end{itemize}

BBHs produce stronger signals and thus achieve higher SNRs, making them more likely to be detected even at large distances. In contrast, BNSs, due to their lower masses, tend to have weaker signals, limiting their detectability to lower redshifts.

\begin{figure}
    \centering
    \includegraphics[width=0.5\textwidth]{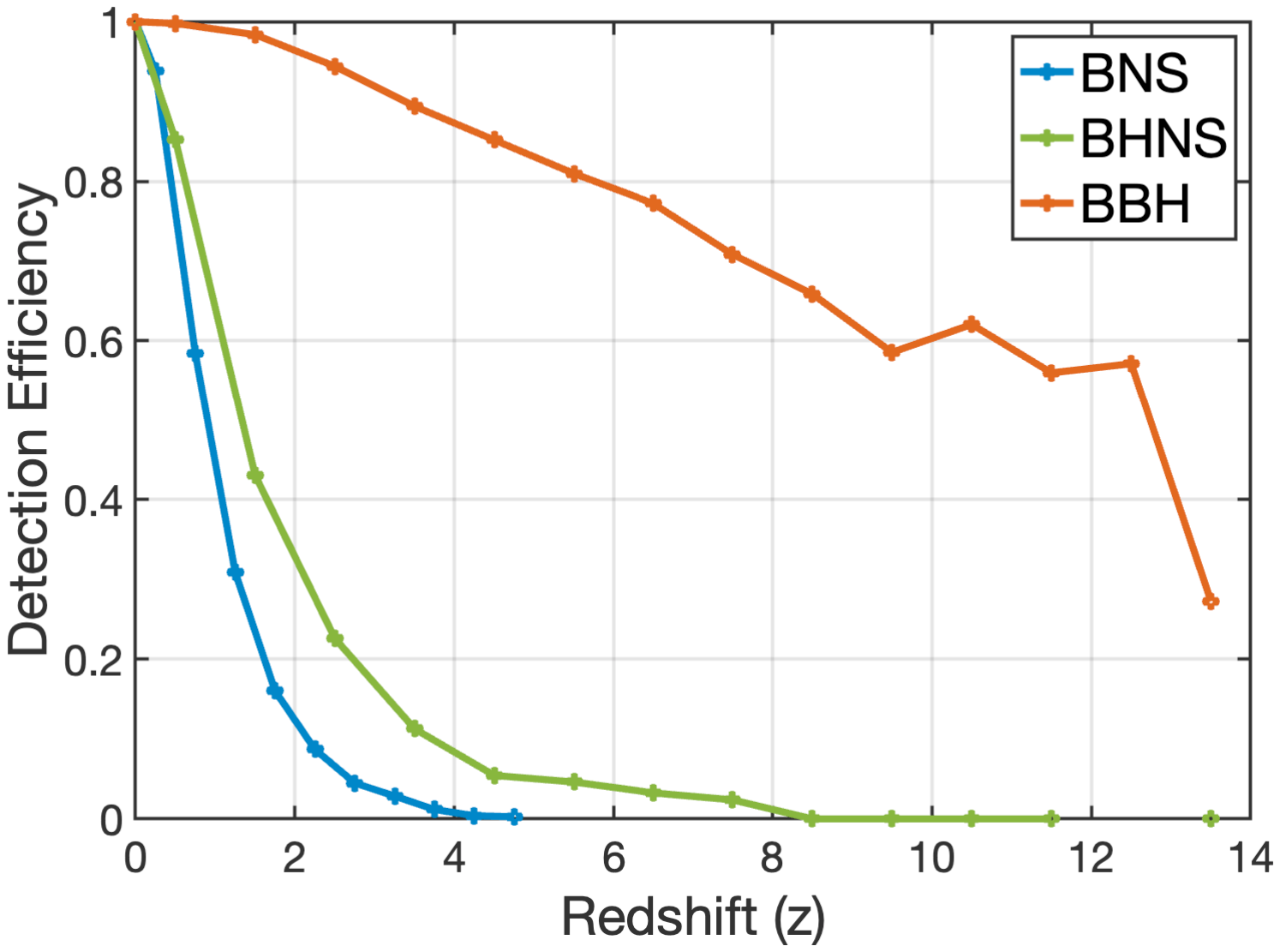}
    \caption{Detection efficiency as a function of redshift for BNSs, BHNSs, and BBHs in our dataset.}
    \label{fig-efficiency}
\end{figure}

To quantify this effect, Fig.~\ref{fig-efficiency} shows the detection efficiency, defined as the fraction of sources detected as a function of redshift for BNSs, BHNSs, and BBHs. At low redshift ($z \approx 0$), nearly all sources are detected. As redshift increases, sources with less favorable orientations or weaker signals drop below the detection threshold.

Note that the detected population density as a function of redshift, $N_{\text{det}}(z)$, is obtained by multiplying the intrinsic population density $N(z)$ by the detection efficiency $\varepsilon(z)$:  
\begin{equation}  
N_{\text{det}}(z) = \varepsilon(z) N(z)  
\end{equation}

\begin{figure}
    \centering
    \includegraphics[width=0.5\textwidth]{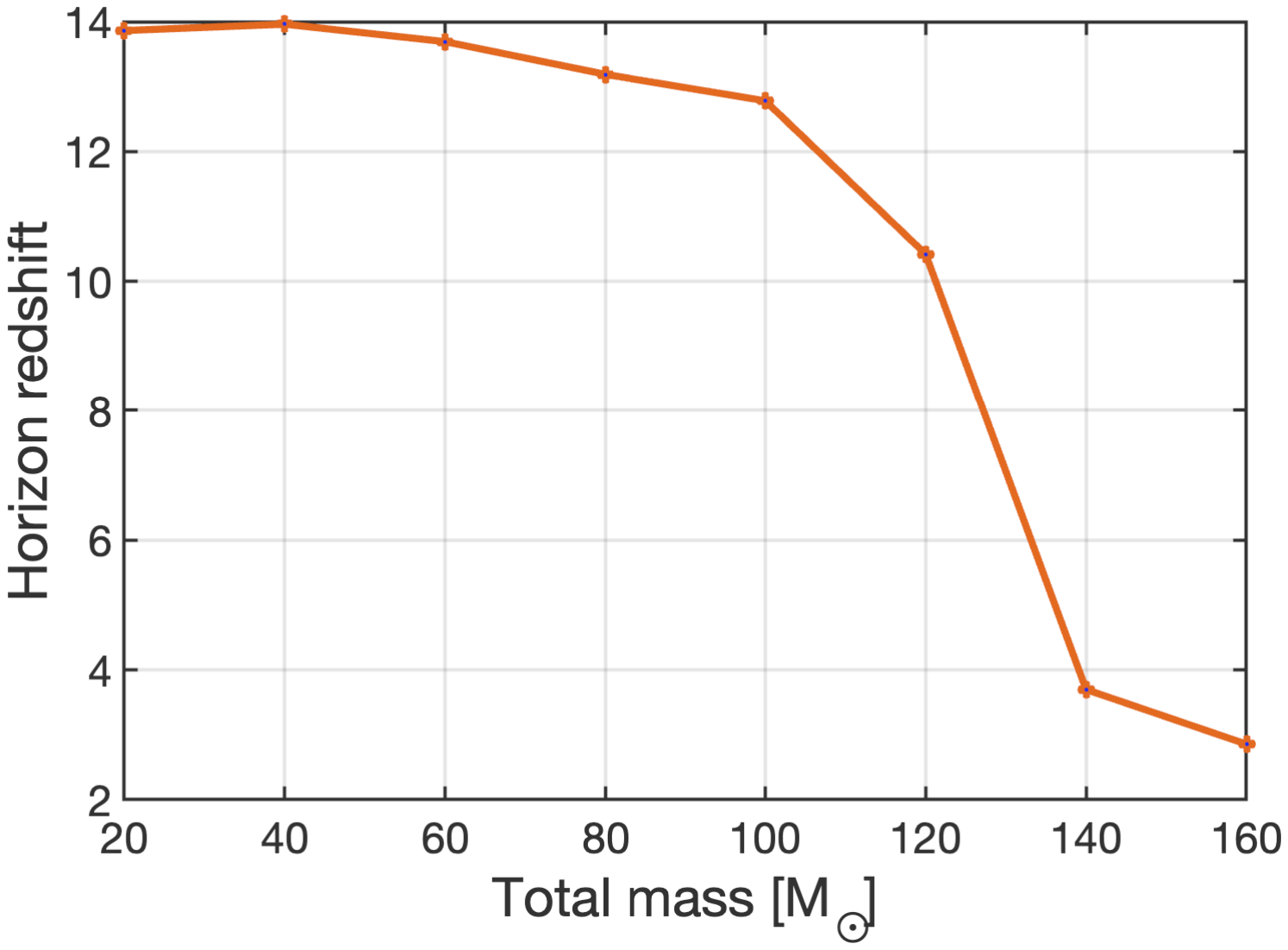}
    \caption{Maximum detected redshift as a function of the observed total mass $M_{z,\text{tot}}$ for the population of BBHs in our dataset.}
    \label{fig-mass-redshift}
\end{figure}

The last redshift before the efficiency curve drops to zero corresponds to the \textit{detection horizon}, beyond which no signals are observed. The detection horizon depends on both mass and orientation: high-mass binaries remain detectable at greater distances due to their stronger signals (see Fig.~\ref{fig-mass-redshift}), while sources with optimal inclination and sky location can be observed further than average. 

While redshift and mass strongly affect detectability, the sky location can also influence the SNR of a signal due to the antenna response pattern of the detector. However, the ET, with its triangular configuration of three co-located interferometers, is designed to provide nearly full-sky coverage, reducing directional sensitivity biases compared to current detectors.

\begin{figure}
    \centering
    \includegraphics[width=\columnwidth]{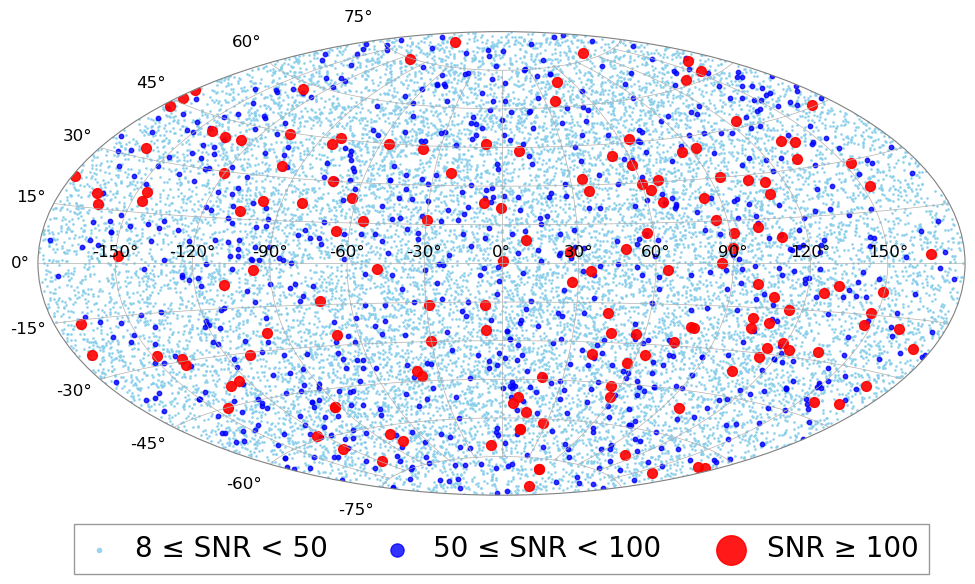}
    \caption{
        Sky map (Aitoff projection in equatorial coordinates) showing detected sources in three signal-to-noise ratio (SNR) categories. 
        Sources with $8 \leq \text{SNR} < 50$ are shown in sky blue with small markers, 
        $50 \leq \text{SNR} < 100$ in blue with medium markers, 
        and $\text{SNR} \geq 100$ in red with large markers. 
        Marker sizes increase with SNR to highlight the strength of detections. 
        The distribution is consistent with an isotropic sky response from the ET network, 
        showing no directional preference or clustering for loud sources.
    }
    \label{fig-skymap-snr}
\end{figure}

To examine potential directional selection effects, we constructed a sky map of all detected sources, grouped into three signal-to-noise ratio (SNR) categories: $8 \leq \text{SNR} < 50$, $50 \leq \text{SNR} < 100$, and $\text{SNR} \geq 100$. As shown in Fig.~\ref{fig-skymap-snr}, the distribution of sources remains isotropic across all SNR levels, with no evidence of clustering or gaps in the sky. This confirms the uniform angular sensitivity of the ET network and suggests no intrinsic bias in detectability based on sky position. High-SNR sources ($\text{SNR} \geq 100$) appear well-distributed, further reinforcing the robustness of the network across the entire celestial sphere.

\begin{figure}
    \centering
    \includegraphics[width=\columnwidth]{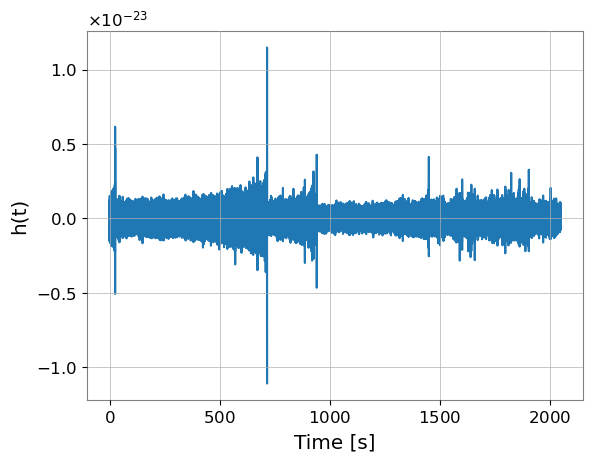}\\
    \includegraphics[width=\columnwidth]{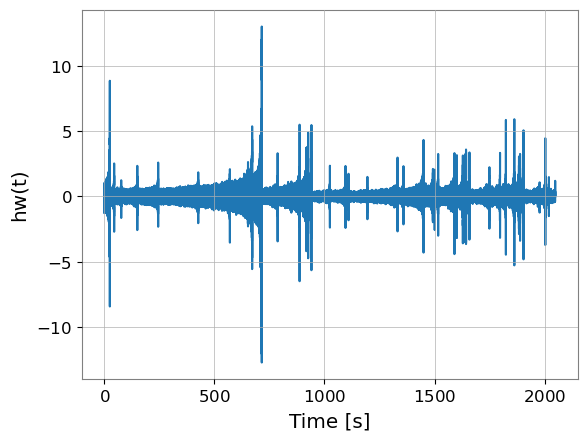}
    \caption{Effect of whitening on a GW signal. The top panel shows the original unwhitened signal, which includes strong low-frequency components. The bottom panel shows the same signal after whitening, where these low-frequency features are suppressed, making individual events more distinguishable. This example corresponds to the first segment for E1, containing only BNS signals.
}
    \label{fig-whitening}
\end{figure}

The detectability presented here assumes no overlap between BNS signals, which could complicate detection. However, whitening significantly attenuates this effect.  
This process removes the frequency-dependent noise structure by dividing the frequency-domain signal \( \tilde{h}(f) \) by the square root of the detector’s noise power spectral density \( S_n(f) \):
\begin{equation}
\tilde{h}_{\text{w}}(f) = \frac{\tilde{h}(f)}{\sqrt{S_n(f)}}
\label{eq-whitening}
\end{equation}
As shown in Fig.~\ref{fig-whitening}, whitening the simulated GW signal helps suppress its low-frequency content and reveals features that might otherwise be masked. While whitening does not resolve the issue of overlapping signals, it helps reduce the dominance of low-frequency components, making individual signals more visually distinguishable, especially when they are close in time or frequency.

\section{Challenges}

For this first ET MDC, we have defined two challenges, one for people new in the field and another for those more experienced in GW data analysis. The challenge is organized by the Data Analysis Division of the Observational Board of the Einstein Telescope and is open to all interested participants. Readers who wish to participate are encouraged to visit the ET webpage (\url{https://www.et-gw.eu/index.php/presentation}) and to subscribe to the OSB Data Analysis mailing list for updates.

\subsection{Challenge 1: Detection of Loudest Sources}

The first challenge consists of finding the six loudest sources with an SNR larger than 300, given the time window. The times and parameters of these sources (5 BBHs and one BNS) are given in Table~\ref{tab:loudest}.

\begin{table*}
    \centering
    \begin{tabular}{|c|c|c|c|c|c|c|c|c|c|}
        \hline
        Type & $t_0$ (GPS) & $t_c$ (GPS) & $\delta1$ (s) & $\delta2$ (s) & $\delta3$ (s)  & $m_1$ (M$_\odot$) & $m_2$ (M$_\odot$) & $z$  & SNR \\ \hline
        BBH   & 1001620399   & 1001620460  & -0.018553 & -0.018553 &-0.018567& 21.9 & 21.5 & 0.11  & 588   \\ \hline
        BBH   & 1001622646   & 1001622670  & -0.012859 & -0.012863 &-0.012884& 35.8  & 32.9 & 0.30  & 387  \\ \hline
        BBH  & 1001560338  & 1001560350 &-0.020699& -0.020695& -0.020702& 44.4 & 44.2 & 0.53 & 379  \\ \hline
        BNS  & 1001330153  & 1001334396  & -0.019703& -0.019708& -0.019695 & 2.3  & 1.3 & 0.042 & 379  \\ \hline
        BBH  & 1002152252 & 1002152286  &0.019105 & 0.019101& 0.019115& 24.8  & 24.3 & 0.43 & 344  \\ \hline
        BBH  & 1000245898 & 1000245939 & 0.012843&0.012860 & 0.012834 & 24.1  & 23.7 & 0.29  & 306  \\ \hline
    \end{tabular}
    \caption{Parameters of the six loudest sources in the dataset (5 BBHs and 1 BNS) used for the first challenge. 
$t_0$ is the time when the source enters the band at 5 Hz, and $t_c$ is the coalescence time, both defined at the Earth's center. 
The columns $\delta1$, $\delta2$, and $\delta3$ give the time delays (in seconds) between the signal arrival at the Earth's center and at detectors E1, E2, and E3, respectively. 
$m_1$ and $m_2$ are the component masses in the source frame, $z$ is the redshift, and the last column shows the combined signal-to-noise ratio (SNR).}
    \label{tab:loudest}
\end{table*}

\begin{figure}[h]
    \centering
    \includegraphics[width=\columnwidth]{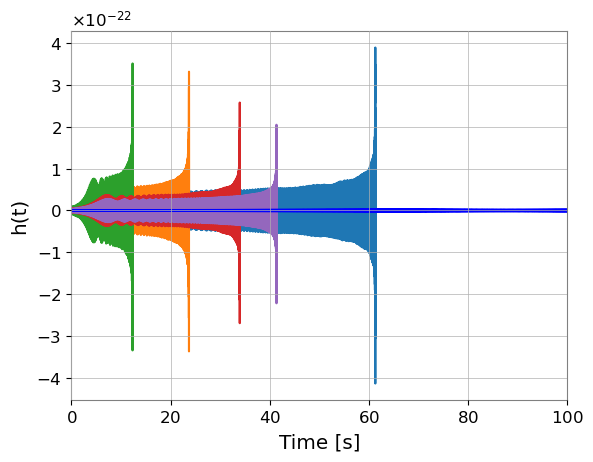}\\
    \includegraphics[width=\columnwidth]{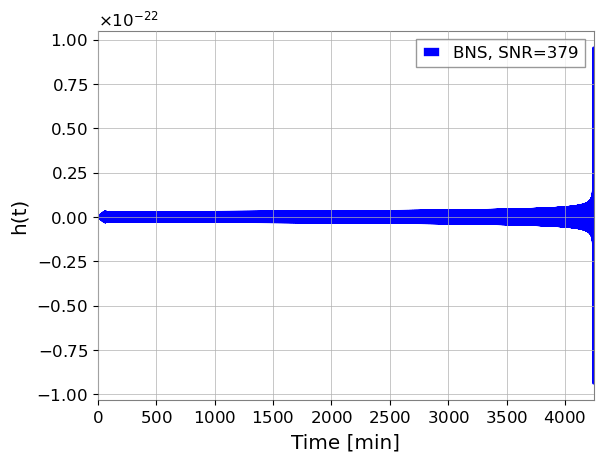}
    \caption{Waveforms of the six loudest sources in the dataset, aligned to a common starting point. The top plot shows the BBH waveforms, while the bottom plot shows the full BNS signal, which lasts significantly longer due to its lower mass.}
    \label{fig-loudest}
\end{figure}

The waveforms of these six loudest sources are shown in Fig.~\ref{fig-loudest}. The main plot displays the waveforms for the five BBH sources, while the inset highlights the BNS waveform, which lasts significantly longer due to its lower mass.

\subsection{Challenge 2: Full Data Set Analysis}

For the second challenge, participants are asked to:
\begin{itemize}[label=\scriptsize$\bullet$]
    \item Find the parameters of the very high SNR sources from the first challenge, the component masses, the component spins for BBHs, the luminosity distance, and the equation of state for the BNS.
    
    \item Perform analysis on the full dataset, which includes dealing with long-duration signals and overlapping sources. The metrics that will be compared are the total number of sources detected and the distribution of the SNR. Participants are also encouraged to extract the distribution of parameters such as the masses, the spins, the luminosity distance, and the redshift.
\end{itemize}

For each MDC round, we aim to produce a joint publication presenting the results and comparing the performance of the different analysis pipelines.

\section{Access to the dataset}

The data for ET-MDC1, for each of the three detectors E1, E2, E3, and the null stream E0, are split into 1300 segments of length 2048s, sampled at 8192 Hz, and are released as frame files, a format typically used in the LIGO-Virgo-Kagra collaboration. It is crucial to emphasize that the signal, including the GW waveforms, as well as the noise, are continuous from one segment to the other.

We provide different sets of data:
\begin{itemize}[label={}] 
    \item \textbf{data\_noise}: Contains only the noise.
    \item \textbf{data\_bbh}: Contains the GW signals for BBHs.
    \item \textbf{data\_bhns}: Contains the GW signals for BHNSs.
    \item \textbf{data\_bns}: Contains the GW signals for BNSs.
    \item \textbf{data}: The final dataset where both noise and signals are combined.
\end{itemize}

In addition, we provide the lists of the source parameters for each of the segments:
\begin{itemize}[label={}] 
    \item \textbf{params\_bbh}: Parameters for BBHs.
    \item \textbf{params\_bhns}: Parameters for BHNSs.
    \item \textbf{params\_bns}: Parameters for BNSs.
\end{itemize}

It is also possible to have the full list of parameters in the files \textit{BBH.txt}, \textit{BHNS.txt}, and \textit{BNS.txt} in the directory \textbf{lists}. The columns in these files include:
\begin{itemize}[label=\scriptsize$\bullet$]
    \item Source number
    \item GPS starting time when entering the band at 5 Hz
    \item GPS time at the maximum of the waveform, associated with the coalescence time
    \item Redshifted mass of the first component
    \item Redshifted mass of the second component
    \item Dimensionless spin of the first component, followed by projections on the x, y, and z axes (only BBHs)
    \item Dimensionless spin of the second component, followed by projections on the x, y, and z axes (only BBHs)
    \item Tidal parameter of the first component (only BNSs)
    \item Tidal parameter of the second component (only BNSs)
    \item Redshift
    \item Luminosity distance in Mpc
    \item Right ascension in radians
    \item Declination in radians
    \item Polarization angle in radians
    \item Inclination in radians
    \item Initial phase in radians
    \item Optimal SNR calculated from Eq.~\ref{eq-snrtot}
    \item Source type: BNS=1, BHNS=2, BBH=3
\end{itemize}

Note that the times are given for the center of the Earth. To obtain the exact times at the detector, a delay (a few or tens of milliseconds) must be applied, which depends on the position of the source and the location of the detectors.

In addition, we provide a training dataset that includes the detector noise along with only the six loudest events. This subset is sufficient for participants tackling the beginner challenge, as it contains all the signals required for the task while allowing them to familiarize themselves with the data format and detection methods before analyzing the full dataset.

Details on accessing the data can be found at the following link\footnote{http://et-origin.cism.ucl.ac.be}.

\section{Simple tutorial}

This tutorial offers a minimal working example to help users get started with the MDC1 dataset. It is intended for those who are new to GW data analysis or looking for a straightforward entry point into working with these data.

We focus on recovering the loudest BBH signal from the catalog using a very simple matched-filtering procedure. The analysis is deliberately basic: we use only a single template that exactly matches the parameters of the injected signal, without exploring a broader parameter space or performing parameter estimation. The idea is to show, step by step, how to access the data, process it, and detect a known signal.

The tutorial guides the reader through reading the injection parameters, loading the relevant data from frame files, visualizing the signal, estimating the noise, and finally applying matched filtering to recover the signal. 
The code and figures in this tutorial are meant to be run interactively in a Jupyter notebook \footnote{https://gitlab.et-gw.eu/osb/div10/mdc-tutorial/MDC\_matched\_filtering.ipynb}, but can easily be adapted for use in scripts or other environments.
Before running this notebook, make sure the necessary data files have been downloaded and unzipped as instructed in the accompanying \texttt{README.md} file.
\subsection*{Importing Libraries}

We begin by importing essential Python libraries for data analysis and plotting. We also define the base directory and subdirectories for the frame and metadata files.

\begin{lstlisting}
import warnings
warnings.filterwarnings("ignore", "Wswiglal-redir-stdio")
import os
import numpy as np
import pandas as pd
import seaborn as sns
import matplotlib.pyplot as plt
import pylab
import gwpy

# Define directories 
base_dir = './'
frames_dir = './frames/'
metadata_dir = './metadata/'
\end{lstlisting}

\subsection*{Read the parameters of injected sources}
We load the injection catalog ordered by decreasing SNR. We select the loudest injection and extract its parameters such as masses, spins, orientation, redshift, etc. The time of signal start and coalescence are also read to estimate the duration.

\begin{lstlisting}
# Read the list of injections ordered by decreasing the signal-to-noise ratio
list_path = os.path.join(metadata_dir, 'list_mdc1_v2.txt')
signals = pd.read_csv(list_path, sep=' ')
i = 0 #loudest injection
pars = signals.iloc[i]
t0 =pars['t0']
tc =pars['tc']
mz1 = pars['mz1']
mz2 = pars['mz2']
chi1 = pars['chi1']
chi1x = pars['chi1x']
chi1y = pars['chi1y']
chi1z = pars['chi1z']
chi2 = pars['chi2']
chi2x = pars['chi2x']
chi2y = pars['chi2y']
chi2z = pars['chi2z']
lambda1 = pars['lambda1']
lambda2 = pars['lambda2']
z = pars['z']
dist = pars['dist']
ra = pars['ra']
decl = pars['decl']
psi = pars['psi']
inc = pars['inc']
phi0 = pars['phi0']
SNR = pars['SNR']
type = pars['type']

deltat=tc-t0  # calculate the duration of the waveform
print(f'Inj {int(pars["#"])},t0{pars["t0"]}, tc {pars["tc"]}, Masses: {pars["mz1"],pars["mz2"]}, deltat {deltat} s, SNR {pars["SNR"]}'

Inj 42581,t01001620399.26659, tc 1001620460.30956, Masses: (24.386689, 23.987984), deltat 61.042969942092896 s, SNR 587.615042
\end{lstlisting}

\subsection*{Read frames with signal only}
\begin{figure}
    \centering
    \includegraphics[width=0.5\textwidth]{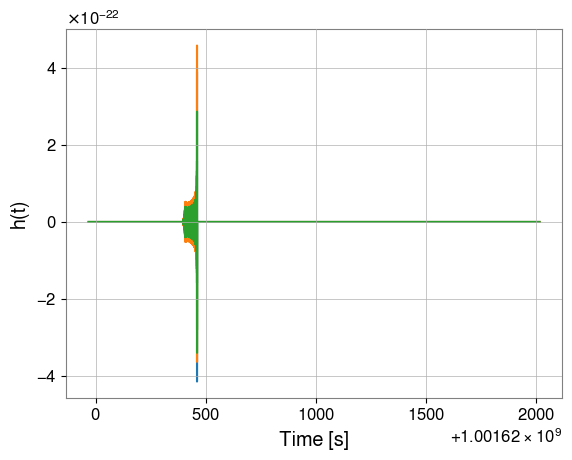}
    \caption{Strain data for signal-only frames from the three ET channels}
    \label{fig-tutorial1}
\end{figure}

We read the GW signal (BBH) without noise from three detectors (E1, E2, E3) (Fig.~\ref{fig-tutorial1}). These frames are stored locally in a compressed format and should be extracted before use.

\begin{lstlisting}
from pycbc import frame

# Read the frames
h1_path = os.path.join(frames_dir,'E-E1_STRAIN_BBH-1001619968-2048.gwf')
h2_path = os.path.join(frames_dir,'E-E2_STRAIN_BBH-1001619968-2048.gwf')
h3_path = os.path.join(frames_dir,'E-E3_STRAIN_BBH-1001619968-2048.gwf')

h1 = frame.read_frame(frameh1_path,'E1:STRAIN')
h2 = frame.read_frame(frameh2_path,'E2:STRAIN')
h3 = frame.read_frame(frameh3_path,'E3:STRAIN')

# Plot the time series
pylab.plot(h1.sample_times,h1)
pylab.plot(h2.sample_times,h2)
pylab.plot(h3.sample_times,h3)
pylab.xlabel('Time (s)')
pylab.ylabel('h(t)')
\end{lstlisting}

\subsection*{Build a simple template}
\begin{figure}
    \centering
    \includegraphics[width=0.45\textwidth]{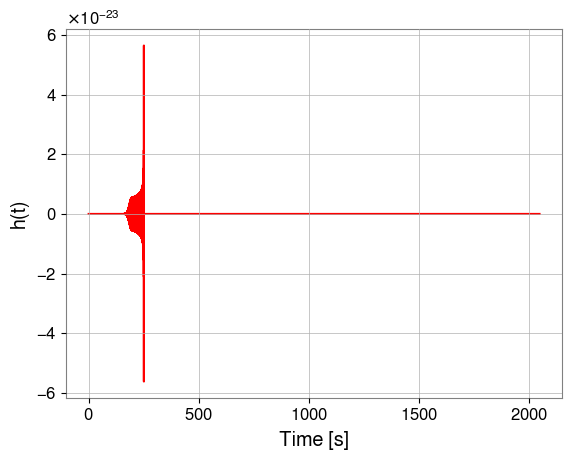}
    \caption{Template based on the true parameters of the source.}
    \label{fig-template}
\end{figure}

We generate a gravitational waveform template in the time domain based on the physical parameters of a compact binary source (Fig.~\ref{fig-template}). Using the \texttt{PyCBC} library~\cite{pycbc}, we compute the gravitational waveform using the IMRPhenomD approximant. The template is constructed using only the 
plus polarization component of the signal, rescaled and aligned for comparison with detector data. For simplicity, we ignore the modulation of the signal due to the source’s motion across the sky, which is justified by the short duration of the signal.

\begin{lstlisting}
from pycbc import types
from pycbc.waveform import get_td_waveform
hp, hc = get_td_waveform(approximant="IMRPhenomD",
                         mass1=mz1,
                         mass2=mz2,
                         spin1z=chi1z,
                         spin2z=chi2z,
                         inclination=inc,
                         delta_t=h1.delta_t,
                         f_lower=5)
template = types.TimeSeries(hp*1e-4, delta_t= h1.delta_t, epoch=0)
template.resize(len(h1))

pylab.plot(template.sample_times,template, color='red')
pylab.xlabel('Time (s)')
pylab.ylabel('h(t)')
\end{lstlisting}

\subsection*{Calculate the PSD for this segment from the null stream}
\begin{figure}
    \centering
    \includegraphics[width=0.45\textwidth]{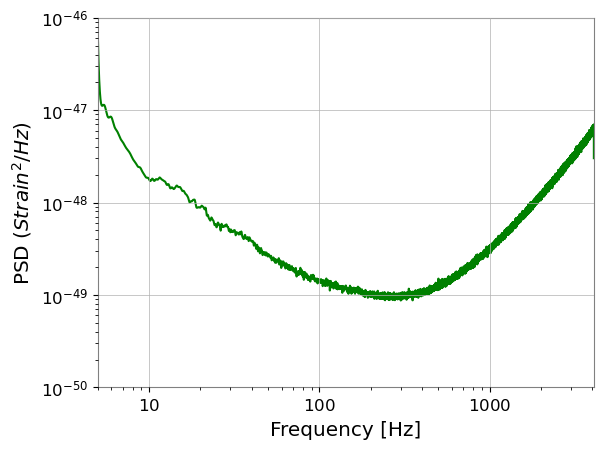}
    \caption{Estimated PSD from the null stream channel}
    \label{fig-tutorial3}
\end{figure}

We read the E0 channel, which includes noise only (Fig.~\ref{fig-tutorial3}). We compute the Power Spectral Density (PSD) using Welch’s method and apply preprocessing steps for whitening.

\begin{lstlisting}

# Read the null stream
frameE0_path = os.path.join(frames_dir,'E-E0_STRAIN-1001619968-2048.gwf')
E0 = frame.read_frame(frameE0_path,'E0:STRAIN')

# Calculate the PSD
psd0 = E0.psd(4)
psd0 = interpolate(psd0, E0.delta_f)
psd0 = inverse_spectrum_truncation(psd0, int(4 * E0.sample_rate), low_frequency_cutoff=5)
pylab.loglog(psd0.sample_frequencies, psd0, color='green')

# Plot the PSD
pylab.ylabel('PSD ($Strain^2 / Hz)$')
pylab.xlabel('Frequency (Hz)')
pylab.xlim(5, 4096)
pylab.ylim(1.e-50, 1.e-46)
\end{lstlisting}

\subsection*{Read the frames with signal+noise}
\begin{figure}
    \centering
    \includegraphics[width=0.5\textwidth]{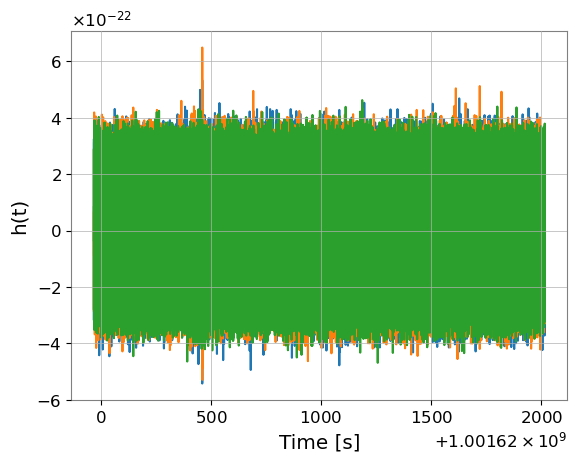}
    \caption{Signal and noise data for each ET channel}
    \label{fig-tutorial4}
\end{figure}

We now read the full data, including both GW signal and simulated noise, for each ET channel (Fig.~\ref{fig-tutorial4}). These are the data we will apply matched filtering to.

\begin{lstlisting}
# Read the frames
E1_path = os.path.join(frames_dir,'E-E1_STRAIN-1001619968-2048.gwf')
E2_path = os.path.join(frames_dir,'E-E2_STRAIN-1001619968-2048.gwf')
E3_path = os.path.join(frames_dir,'E-E3_STRAIN-1001619968-2048.gwf')

E1 = frame.read_frame(E1_path,'E1:STRAIN')
E2 = frame.read_frame(E2_path,'E2:STRAIN')
E3 = frame.read_frame(E3_path,'E3:STRAIN')

# Read the time series
pylab.plot(E1.sample_times,E1)
pylab.plot(E2.sample_times,E2)
pylab.plot(E3.sample_times,E3)
pylab.xlabel('Time (s)')
pylab.ylabel('h(t)')
\end{lstlisting}

\subsection*{Plot the Spectrogram}
\begin{figure}
    \centering
    \includegraphics[width=0.5\textwidth]{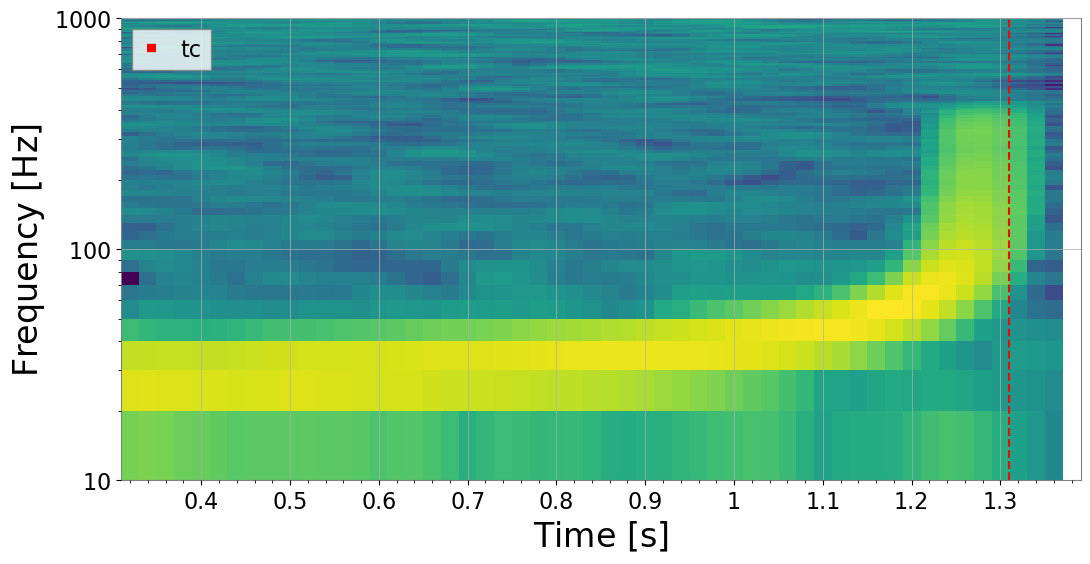}
    \caption{Spectrogram showing signal around the coalescence time}
    \label{fig-tutorial5}
\end{figure}

We compute and plot a spectrogram around the coalescence time using the E1 data, transformed to GWpy. The coalescence is marked with a vertical dashed line (Fig.~\ref{fig-tutorial5}).

\begin{lstlisting}
from gwpy.timeseries import TimeSeries as GWpyTimeSeries

# Convert PyCBC TimeSeries (E1) into GWpy
gwpy_E1 = GWpyTimeSeries(E1.numpy(), sample_rate=E1.sample_rate, t0=float(E1.start_time))

# Crop around coalescence
gwpy_E1_crop = gwpy_E1.crop(tc - 1, tc + 0.1)

# Compute spectrogram on cropped data 
sg = gwpy_E1_crop.spectrogram2(fftlength=0.1, overlap=0.08)**0.5

# Plot it
fig = sg.plot(norm='log', yscale='log', ylim=(10, 1000))
fig.gca().axvline(tc, color='red', linestyle='--', label='tc')
fig.gca().legend()
\end{lstlisting}

\subsection*{Apply match filtering}
\begin{figure}
    \centering
    \includegraphics[width=0.5\textwidth]{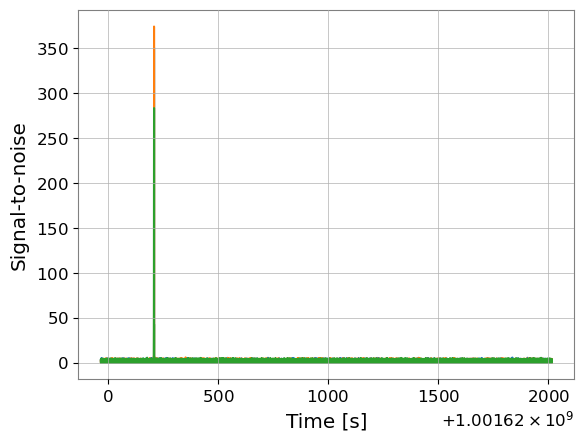}
    \caption{SNR time series from matched filtering for each ET channel}
    \label{fig-tutorial6}
\end{figure}

We filter the signal+noise data using a known template and the estimated PSD. Peaks in the matched filter SNR time series indicate likely signal events (Fig.~\ref{fig-tutorial6}).

\begin{lstlisting}
from pycbc.filter import matched_filter

snr1 = matched_filter(template, E1, psd=psd0, low_frequency_cutoff=5)
snr2 = matched_filter(template, E2, psd=psd0, low_frequency_cutoff=5)
snr3 = matched_filter(template, E3, psd=psd0, low_frequency_cutoff=5)

pylab.figure(figsize=[10, 4])
pylab.plot(snr1.sample_times, abs(snr1))
pylab.plot(snr2.sample_times, abs(snr2))
pylab.plot(snr3.sample_times, abs(snr3))
pylab.ylabel('Signal-to-noise')
pylab.xlabel('Time (s)')
pylab.show()

# Find peaks
from scipy.signal import find_peaks

peak1 = abs(snr1).numpy().argmax()
snrp1 = snr1[peak1]
time1 = snr1.sample_times[peak1]
print("We found a signal at {}s with SNR {} in E1".format(time1, abs(snrp1)))

peak2 = abs(snr2).numpy().argmax()
snrp2 = snr2[peak2]
time2 = snr2.sample_times[peak2]
print("We found a signal at {}s with SNR {} in E2".format(time2, abs(snrp2)))

peak3 = abs(snr3).numpy().argmax()
snrp3 = snr3[peak3]
time3 = snr3.sample_times[peak3]
print("We found a signal at {}s with SNR {} in E3".format(time3, abs(snrp3)))

We found a signal at 1001620209.295044s with SNR 350.671256381459 in E1
We found a signal at 1001620209.295044s with SNR 374.21651082700436 in E2
We found a signal at 1001620209.295044s with SNR 283.38684302783685 in E3

\end{lstlisting}

\section{Conclusion}

The ET community has begun a series of MDCs to prepare for the operational complexities of the upcoming third-generation GW detectors.
The ET mock data challenges aim to identify the limitations of current data analysis methods and guide the development of new approaches capable of operating in the complex, signal-rich environments anticipated for third-generation detectors. These challenges also provide a platform for validating different detector configurations and assessing whether data analysis frameworks can scale to meet future computational demands. Bayesian methods have shown strong potential for parameter estimation in GW astronomy, but their application to long-duration and overlapping signals remains a developing area. Machine learning is expected to play an increasingly important role in this context, offering novel strategies for enhancing signal detection, managing signal overlaps, and enabling faster and more robust parameter estimation. For a review of current strategies, see \cite{2025arXiv250312263A}.

This paper introduces the first round of the MDC for the ET, offering a comprehensive overview of the dataset generation, statistical analysis, and the properties of the GW signals.

To ensure a controlled setting for this initial round, the data includes only CBC signals embedded in stationary Gaussian noise. This simplification facilitates a systematic evaluation of existing detection pipelines and parameter estimation methods, while also providing a manageable framework for the development of new analysis strategies.

Two types of challenges are presented: a beginner challenge focused on detecting the six loudest events, and an expert challenge requiring the analysis of the full dataset, including long-duration and overlapping signals.

A first publication based on this dataset, focusing on unmodeled search methods, has already been released \cite{2025PhRvD.111b2002M}. In that work, the PySTAMPAS pipeline was applied to the MDC dataset, successfully recovering a large fraction of the injected CBC signals despite the absence of templates. This demonstrates the potential of unmodeled approaches for low-latency analyses and as a complement to matched-filter pipelines.

Future rounds of the MDC will introduce more complexity, incorporating realistic noise features such as glitches, gaps, and correlated noise, as well as a broader variety of source types, including bursts, continuous waves, and stochastic backgrounds. We also plan to investigate different detector configurations beyond the triangular ET layout, such as networks of separated L-shaped detectors with varying arm lengths.

For each round, we aim to produce a joint publication comparing the performance of various pipelines and methodologies. The results will contribute to refining detection pipelines, parameter estimation techniques, and computational strategies, offering valuable insights for the development of more efficient methods. As the complexity of the mock data increases, the knowledge gained will ensure that the community is well-prepared for both the astrophysical discoveries and the substantial computational demands that the ET will impose.

\section*{Acknowledgments}
We acknowledge Andres Tanasijczuk for his technical support in making the frames publicly accessible and for documenting the access procedures. 
We gratefully acknowledge John Veitch for his tutorial on the MDC1 dataset, which provided the essential foundation for the tutorial presented in this manuscript.
We thank Michael Ebersold and Gianluca Inguglia for insightful comments in the early version of the manuscript, and also Markus Bachlechner and Adrian Macquet for their critical feedback which helped us identify and correct issues in an early version of the ET MDC dataset. We thank the Einstein Telescope Data Analysis division members, especially the chairs Elena Cuocco, Gianluca Guidi, and Anuradha Samajar. 

\bibliography{references}

@article{LISAMDC,
    author = "Babak, Stanislav and others",
    editor = "Marka, Zsuzsa and Marka, Szabolcs",
    collaboration = "Mock LISA Data Challenge Task Force",
    title = "{The Mock LISA Data Challenges: From Challenge 3 to Challenge 4}",
    eprint = "0912.0548",
    archivePrefix = "arXiv",
    primaryClass = "gr-qc",
    doi = "10.1088/0264-9381/27/8/084009",
    journal = "Class. Quant. Grav.",
    volume = "27",
    pages = "084009",
    year = "2010"
}

@ARTICLE{2007CQGra..24S.551A,
       author = {{Arnaud}, K.~A. and {Babak}, S. and {Baker}, J.~G. and {Benacquista}, M.~J. and {Cornish}, N.~J. and {Cutler}, C. and {Finn}, L.~S. and {Larson}, S.~L. and {Littenberg}, T. and {Porter}, E.~K. and {Vallisneri}, M. and {Vecchio}, A. and {Vinet}, J. -Y. and {Data Challenge Task Force}, The Mock LISA},
        title = "{An overview of the second round of the Mock LISA Data Challenges}",
      journal = {Classical and Quantum Gravity},
         year = 2007,
        month = oct,
       volume = {24},
       number = {19},
        pages = {S551-S564},
          doi = {10.1088/0264-9381/24/19/S18},
archivePrefix = {arXiv},
       eprint = {gr-qc/0701170},
 primaryClass = {gr-qc},
       adsurl = {https://ui.adsabs.harvard.edu/abs/2007CQGra..24S.551A}
}

@ARTICLE{2022arXiv220412142B,
       author = {{Baghi}, Quentin},
        title = "{The LISA Data Challenges}",
      journal = {arXiv e-prints},
         year = 2022,
        month = apr,
          eid = {arXiv:2204.12142},
        pages = {arXiv:2204.12142},
          doi = {10.48550/arXiv.2204.12142},
archivePrefix = {arXiv},
       eprint = {2204.12142},
 primaryClass = {gr-qc},
       adsurl = {https://ui.adsabs.harvard.edu/abs/2022arXiv220412142B}
}

@ARTICLE{2012PhRvD..86l2001R,
       author = {{Regimbau}, Tania and {Dent}, Thomas and {Del Pozzo}, Walter and {Giampanis}, Stefanos and {Li}, Tjonnie G.~F. and {Robinson}, Craig and {Van Den Broeck}, Chris and {Meacher}, Duncan and {Rodriguez}, Carl and {Sathyaprakash}, B.~S. and {W{\'o}jcik}, Katarzyna},
        title = "{Mock data challenge for the Einstein Gravitational-Wave Telescope}",
      journal = {\prd},
         year = {2012},
        month = {dec},
       volume = {86},
       number = {12},
          eid = {122001},
        pages = {122001},
          doi = {10.1103/PhysRevD.86.122001},
archivePrefix = {arXiv},
       eprint = {1201.3563},
 primaryClass = {gr-qc},
       adsurl = {https://ui.adsabs.harvard.edu/abs/2012PhRvD..86l2001R}
}

@ARTICLE{2014PhRvD..89h4046R,
       author = {{Regimbau}, Tania and {Meacher}, Duncan and {Coughlin}, Michael},
        title = "{Second Einstein Telescope mock science challenge: Detection of the gravitational-wave stochastic background from compact binary coalescences}",
      journal = {\prd},
         year = 2014,
        month = apr,
       volume = {89},
       number = {8},
          eid = {084046},
        pages = {084046},
          doi = {10.1103/PhysRevD.89.084046},
archivePrefix = {arXiv},
       eprint = {1404.1134},
 primaryClass = {astro-ph.CO},
       adsurl = {https://ui.adsabs.harvard.edu/abs/2014PhRvD..89h4046R}
}

@ARTICLE{2016PhRvD..93b4018M,
       author = {{Meacher}, Duncan and {Cannon}, Kipp and {Hanna}, Chad and {Regimbau}, Tania and {Sathyaprakash}, B.~S.},
        title = "{Second Einstein Telescope mock data and science challenge: Low frequency binary neutron star data analysis}",
      journal = {\prd},
         year = 2016,
        month = jan,
       volume = {93},
       number = {2},
          eid = {024018},
        pages = {024018},
          doi = {10.1103/PhysRevD.93.024018},
archivePrefix = {arXiv},
       eprint = {1511.01592},
 primaryClass = {gr-qc},
       adsurl = {https://ui.adsabs.harvard.edu/abs/2016PhRvD..93b4018M}
}

@ARTICLE{2015PhRvD..92f3002M,
       author = {{Meacher}, Duncan and {Coughlin}, Michael and {Morris}, Sean and {Regimbau}, Tania and {Christensen}, Nelson and {Kandhasamy}, Shivaraj and {Mandic}, Vuk and {Romano}, Joseph D. and {Thrane}, Eric},
        title = "{Mock data and science challenge for detecting an astrophysical stochastic gravitational-wave background with Advanced LIGO and Advanced Virgo}",
      journal = {\prd},
         year = 2015,
        month = sep,
       volume = {92},
       number = {6},
          eid = {063002},
        pages = {063002},
          doi = {10.1103/PhysRevD.92.063002},
archivePrefix = {arXiv},
       eprint = {1506.06744},
 primaryClass = {astro-ph.HE},
       adsurl = {https://ui.adsabs.harvard.edu/abs/2015PhRvD..92f3002M}
}

@ARTICLE{2021MNRAS.505..339M,
       author = {{Mapelli}, Michela and {Dall'Amico}, Marco and {Bouffanais}, Yann and {Giacobbo}, Nicola and {Arca Sedda}, Manuel and {Artale}, M. Celeste and {Ballone}, Alessandro and {Di Carlo}, Ugo N. and {Iorio}, Giuliano and {Santoliquido}, Filippo and {Torniamenti}, Stefano},
        title = "{Hierarchical black hole mergers in young, globular and nuclear star clusters: the effect of metallicity, spin and cluster properties}",
      journal = {\mnras},
         year = 2021,
        month = jul,
       volume = {505},
       number = {1},
        pages = {339-358},
          doi = {10.1093/mnras/stab1334},
archivePrefix = {arXiv},
       eprint = {2103.05016},
 primaryClass = {astro-ph.HE},
       adsurl = {https://ui.adsabs.harvard.edu/abs/2021MNRAS.505..339M}
}

@ARTICLE{2022MNRAS.511.5797M,
       author = {{Mapelli}, Michela and {Bouffanais}, Yann and {Santoliquido}, Filippo and {Arca Sedda}, Manuel and {Artale}, M. Celeste},
        title = "{The cosmic evolution of binary black holes in young, globular, and nuclear star clusters: rates, masses, spins, and mixing fractions}",
      journal = {\mnras},
         year = 2022,
        month = apr,
       volume = {511},
       number = {4},
        pages = {5797-5816},
          doi = {10.1093/mnras/stac422},
archivePrefix = {arXiv},
       eprint = {2109.06222},
 primaryClass = {astro-ph.HE},
       adsurl = {https://ui.adsabs.harvard.edu/abs/2022MNRAS.511.5797M}
}

@ARTICLE{2021MNRAS.502.4877S,
       author = {{Santoliquido}, Filippo and {Mapelli}, Michela and {Giacobbo}, Nicola and {Bouffanais}, Yann and {Artale}, M. Celeste},
        title = "{The cosmic merger rate density of compact objects: impact of star formation, metallicity, initial mass function, and binary evolution}",
      journal = {\mnras},
         year = 2021,
        month = apr,
       volume = {502},
       number = {4},
        pages = {4877-4889},
          doi = {10.1093/mnras/stab280},
archivePrefix = {arXiv},
       eprint = {2009.03911},
 primaryClass = {astro-ph.HE},
       adsurl = {https://ui.adsabs.harvard.edu/abs/2021MNRAS.502.4877S}
}

@ARTICLE{2020JCAP...03..050M,
       author = {{Maggiore}, Michele and {Van Den Broeck}, Chris and {Bartolo}, Nicola and {Belgacem}, Enis and {Bertacca}, Daniele and {Bizouard}, Marie Anne and {Branchesi}, Marica and {Clesse}, Sebastien and {Foffa}, Stefano and {Garc{\'\i}a-Bellido}, Juan and {Grimm}, Stefan and {Harms}, Jan and {Hinderer}, Tanja and {Matarrese}, Sabino and {Palomba}, Cristiano and {Peloso}, Marco and {Ricciardone}, Angelo and {Sakellariadou}, Mairi},
        title = "{Science case for the Einstein telescope}",
      journal = {\jcap},
         year = 2020,
        month = mar,
       volume = {2020},
       number = {3},
          eid = {050},
        pages = {050},
          doi = {10.1088/1475-7516/2020/03/050},
archivePrefix = {arXiv},
       eprint = {1912.02622},
 primaryClass = {astro-ph.CO},
       adsurl = {https://ui.adsabs.harvard.edu/abs/2020JCAP...03..050M}
}

@article{2016A&A...594A..13P,
    author = "Ade, P. A. R. and others",
    collaboration = "Planck",
    title = "{Planck 2015 results. XIII. Cosmological parameters}",
    eprint = "1502.01589",
    archivePrefix = "arXiv",
    primaryClass = "astro-ph.CO",
    doi = "10.1051/0004-6361/201525830",
    journal = "Astron. Astrophys.",
    volume = "594",
    pages = "A13",
    year = "2016"
}

@ARTICLE{2010CQGra..27a5003H,
       author = {{Hild}, S. and {Chelkowski}, S. and {Freise}, A. and {Franc}, J. and {Morgado}, N. and {Flaminio}, R. and {DeSalvo}, R.},
        title = "{A xylophone configuration for a third-generation gravitational wave detector}",
      journal = {Classical and Quantum Gravity},
         year = {2010},
        month = {jan},
       volume = {27},
       number = {1},
          eid = {015003},
        pages = {015003},
          doi = {10.1088/0264-9381/27/1/015003},
archivePrefix = {arXiv},
       eprint = {0906.2655},
 primaryClass = {gr-qc},
       adsurl = {https://ui.adsabs.harvard.edu/abs/2010CQGra..27a5003H}
}

@misc{lalsuite,
       author         = "{LIGO Scientific Collaboration} and {Virgo Collaboration} and {KAGRA Collaboration}",
       title          = "{LVK} {A}lgorithm {L}ibrary - {LALS}uite",
       howpublished   = "Free software (GPL)",
       doi            = "10.7935/GT1W-FZ16",
       year           = "2018"
 }

@ARTICLE{2023EPJP..138..352J,
       author = {{Janssens}, Kamiel and {Boileau}, Guillaume and {Bizouard}, Marie-Anne and {Christensen}, Nelson and {Regimbau}, Tania and {Remortel}, Nick van},
        title = "{Formalism for power spectral density estimation for non-identical and correlated noise using the null channel in Einstein Telescope}",
      journal = {European Physical Journal Plus},
         year = 2023,
        month = apr,
       volume = {138},
       number = {4},
          eid = {352},
        pages = {352},
          doi = {10.1140/epjp/s13360-023-03948-9},
archivePrefix = {arXiv},
       eprint = {2205.00416},
 primaryClass = {gr-qc},
       adsurl = {https://ui.adsabs.harvard.edu/abs/2023EPJP..138..352J}
}

@ARTICLE{2022PhRvD.106d2008J,
       author = {{Janssens}, Kamiel and {Boileau}, Guillaume and {Christensen}, Nelson and {Badaracco}, Francesca and {van Remortel}, Nick},
        title = "{Impact of correlated seismic and correlated Newtonian noise on the Einstein Telescope}",
      journal = {\prd},
         year = 2022,
        month = aug,
       volume = {106},
       number = {4},
          eid = {042008},
        pages = {042008},
          doi = {10.1103/PhysRevD.106.042008},
archivePrefix = {arXiv},
       eprint = {2206.06809},
 primaryClass = {astro-ph.IM},
       adsurl = {https://ui.adsabs.harvard.edu/abs/2022PhRvD.106d2008J}
}

@ARTICLE{2023arXiv230502694J,
       author = {{Janssens}, Kamiel},
        title = "{Prospects for an isotropic gravitational wave background detection with Earth-based interferometric detectors and the threat of correlated noise}",
      journal = {arXiv e-prints},
         year = 2023,
        month = may,
          eid = {arXiv:2305.02694},
        pages = {arXiv:2305.02694},
          doi = {10.48550/arXiv.2305.02694},
archivePrefix = {arXiv},
       eprint = {2305.02694},
 primaryClass = {gr-qc},
       adsurl = {https://ui.adsabs.harvard.edu/abs/2023arXiv230502694J}
}

@ARTICLE{2024PhRvD.109j2002J,
       author = {{Janssens}, Kamiel and {Boileau}, Guillaume and {Christensen}, Nelson and {van Remortel}, Nick and {Badaracco}, Francesca and {Canuel}, Benjamin and {Cardini}, Alessandro and {Contu}, Andrea and {Coughlin}, Michael W. and {Decitre}, Jean-Baptiste and {De Rosa}, Rosario and {Di Giovanni}, Matteo and {D'Urso}, Domenico and {Gaffet}, St{\'e}phane and {Giunchi}, Carlo and {Harms}, Jan and {Koley}, Soumen and {Mangano}, Valentina and {Naticchioni}, Luca and {Olivieri}, Marco and {Paoletti}, Federico and {Rozza}, Davide and {Sabulsky}, Dylan O. and {Shani-Kadmiel}, Shahar and {Trozzo}, Lucia},
        title = "{Correlated 0.01-40 Hz seismic and Newtonian noise and its impact on future gravitational-wave detectors}",
      journal = {\prd},
         year = 2024,
        month = may,
       volume = {109},
       number = {10},
          eid = {102002},
        pages = {102002},
          doi = {10.1103/PhysRevD.109.102002},
archivePrefix = {arXiv},
       eprint = {2402.17320},
 primaryClass = {gr-qc},
       adsurl = {https://ui.adsabs.harvard.edu/abs/2024PhRvD.109j2002J}
}

@ARTICLE{2022PhRvD.105l2007G,
       author = {{Goncharov}, Boris and {Nitz}, Alexander H. and {Harms}, Jan},
        title = "{Utilizing the null stream of the Einstein Telescope}",
      journal = {\prd},
         year = 2022,
        month = jun,
       volume = {105},
       number = {12},
          eid = {122007},
        pages = {122007},
          doi = {10.1103/PhysRevD.105.122007},
archivePrefix = {arXiv},
       eprint = {2204.08533},
 primaryClass = {gr-qc},
       adsurl = {https://ui.adsabs.harvard.edu/abs/2022PhRvD.105l2007G}
}

@ARTICLE{2025PhRvD.111b2002M,
       author = {{Macquet}, Adrian and {Dal Canton}, Tito and {Regimbau}, Tania},
        title = "{Weakly modeled search for compact binary coalescences in the Einstein Telescope}",
      journal = {\prd},
         year = 2025,
        month = jan,
       volume = {111},
       number = {2},
          eid = {022002},
        pages = {022002},
          doi = {10.1103/PhysRevD.111.022002},
archivePrefix = {arXiv},
       eprint = {2408.13007},
 primaryClass = {gr-qc},
       adsurl = {https://ui.adsabs.harvard.edu/abs/2025PhRvD.111b2002M}
}

@misc{2025arXiv250312263A,
    author = "Abac, Adrian and others",
    title = "{The Science of the Einstein Telescope}",
    eprint = "2503.12263",
    archivePrefix = "arXiv",
    primaryClass = "gr-qc",
    reportNumber = "ET-0036C-25",
    month = "3",
    year = "2025"
}

@article{LIGOScientific:2016aoc,
    author = "Abbott, B. P. and others",
    collaboration = "LIGO Scientific, Virgo",
    title = "{Observation of Gravitational Waves from a Binary Black Hole Merger}",
    eprint = "1602.03837",
    archivePrefix = "arXiv",
    primaryClass = "gr-qc",
    reportNumber = "LIGO-P150914",
    doi = "10.1103/PhysRevLett.116.061102",
    journal = "Phys. Rev. Lett.",
    volume = "116",
    number = "6",
    pages = "061102",
    year = "2016"
}

@article{pycbc,
    author = "Usman, Samantha A. and others",
    title = "{The PyCBC search for gravitational waves from compact binary coalescence}",
    eprint = "1508.02357",
    archivePrefix = "arXiv",
    primaryClass = "gr-qc",
    reportNumber = "LIGO-P1500086",
    doi = "10.1088/0264-9381/33/21/215004",
    journal = "Class. Quant. Grav.",
    volume = "33",
    number = "21",
    pages = "215004",
    year = "2016"
}

@techreport{T1400308,
  author       = {Duncan Meacher},
  title        = {Einstein Telescope pre site selection detector locations (2014)},
  institution  = {LIGO Document Control Center},
  year         = {2014},
  number       = {LIGO-T1400308},
  note         = {\url{https://dcc.ligo.org/LIGO-T1400308}},
}

@ARTICLE{Madau2017,
       author = {{Madau}, Piero and {Fragos}, Tassos},
        title = "{Radiation Backgrounds at Cosmic Dawn: X-Rays from Compact Binaries}",
      journal = {\apj},
         year = 2017,
        month = may,
       volume = {840},
       number = {1},
          eid = {39},
        pages = {39},
          doi = {10.3847/1538-4357/aa6af9},
archivePrefix = {arXiv},
       eprint = {1606.07887},
 primaryClass = {astro-ph.GA},
       adsurl = {https://ui.adsabs.harvard.edu/abs/2017ApJ...840...39M}
}
\end{document}